\documentclass[fleqn,usenatbib,useAMS]{mnras}

\usepackage{newtxtext,newtxmath}

\usepackage[T1]{fontenc}

\DeclareRobustCommand{\VAN}[3]{#2}
\let\VANthebibliography\thebibliography
\def\thebibliography{\DeclareRobustCommand{\VAN}[3]{##3}\VANthebibliography}

\usepackage{graphicx}	
\usepackage{amsmath}	
\usepackage{scalefnt}
\usepackage{newtxtext,newtxmath}
\usepackage{verbatim}
\usepackage{longtable}
\usepackage{natbib}
\usepackage{multirow}
\usepackage{lscape} 
\usepackage{tabularx}

\usepackage{color,soul}

\newcommand{\orcid}[1]{\href{https://orcid.org/#1}{\includegraphics[width=10pt]{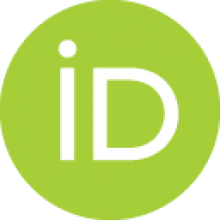}}}

\newbox\grsign \setbox\grsign=\hbox{$>$} \newdimen\grdimen \grdimen=\ht\grsign
\newbox\simlessbox \newbox\simgreatbox
\setbox\simgreatbox=\hbox{\raise.5ex\hbox{$>$}\llap
     {\lower.5ex\hbox{$\sim$}}}\ht1=\grdimen\dp1=0pt
\setbox\simlessbox=\hbox{\raise.5ex\hbox{$<$}\llap
     {\lower.5ex\hbox{$\sim$}}}\ht2=\grdimen\dp2=0pt

\def\simless{\mathrel{\copy\simlessbox}}

\title[Abundances of 58 spheroid bulge stars ]
{Abundance analysis of APOGEE spectra for 58 metal-poor stars from the bulge spheroid}

\author[R. Razera et al.]{
R. Razera,$^{1}$\thanks{E-mail: roberta.razera@usp.br}\orcid{0000-0002-1734-7540}
B. Barbuy,$^{1}$\orcid{0000-0001-9264-4417},
T. C. Moura,$^{1}$\orcid{0000-0003-1548-7221},
H. Ernandes,$^{1}$\orcid{0000-0001-6541-1933},
A. P\'erez-Villegas,$^{2}$\orcid{0000-0002-5974-3998},
S. O. Souza,$^{1,3}$\orcid{0000-0001-8052-969X},
\newauthor
C. Chiappini,$^{3}$\orcid{0000-0003-1269-7282},
A. B. A. Queiroz,$^{3,4}$\orcid{0000-0001-9209-7599},
F. Anders,$^{5}$\orcid{0000-0003-4524-9363},
J. G. Fern\'andez-Trincado,$^{6}$\orcid{0000-0003-3526-5052},
A. C. S. Fria\c ca,$^1$, 
\newauthor
K. Cunha,$^{7,8}$\orcid{0000-0001-9282-715X},
V. V. Smith,$^{9}$\orcid{0000-0002-0134-2024},
B.X. Santiago,$^{10}$\orcid{0000-0001-7692-8495},
R.P. Schiavon,$^{11}$\orcid{0000-0002-2244-0897},
M. Valentini,$^{3}$\orcid{0000-0003-0974-4148},
D. Minniti,$^{12,13}$\orcid{0000-0002-7064-099X},
\newauthor
M. Schultheis,$^{14}$\orcid{0000-0002-6590-1657},
D. Geisler,$^{15,16,17}$\orcid{0000-0002-3900-8208},
J. Sobeck,$^{18}$\orcid{0000-0002-4989-0353},
V. M. Placco,$^{9}$\orcid{0000-0003-4479-1265},
M. Zoccali,$^{19,20}$\orcid{0000-0002-5829-2267}
\\
\\
$^{1}$ Universidade de S\~ao Paulo, IAG, Rua do Mat\~ao 1226, Cidade Universit\'aria, S\~ao Paulo 05508-900, Brazil\\
$^{2}$ Instituto de Astronom\'ia, Universidad Nacional Aut\'onoma de M\'exico, A. P. 106, C.P. 22800, Ensenada, B. C., M\'exico\\
$^{3}$ Leibniz-Institut f\"ur Astrophysik Potsdam (AIP), An der Sternwarte 16, Potsdam, 14482, Germany\\
$^{4}$Institut f\"{u}r Physik und Astronomie, Universit\"{a}t Potsdam, Haus 28 Karl-Liebknecht-Str. 24/25, D-14476 Golm (Potsdam), Germany\\
$^{5}$ Institut de Ci\`encies del Cosmos, Universitat de Barcelona (IEEC-UB), Mart\'{\i} i Franqu\`es 1, 08028 Barcelona, Spain\\
$^{6}$ Instituto de Astronom\'{\i}a, Universidad Cat\'olica del Norte, Av. Angamos 0610, Antofagasta, Chile \\
$^{7}$ University of Arizona, Tucson, AZ 85719, USA \\
$^{8}$ Observat\'orio Nacional, S\ Crist\'ov\~ao, Rio de Janeiro, Brazil\\
$^{9}$ NSF’s NOIRLab, 950 N. Cherry Ave., Tucson, AZ 85719, USA \\
$^{10}$ Universidade Federal do Rio Grande do Sul, Caixa Postal 15051, 91501-970 Porto Alegre, Brazil \\
$^{11}$ Astrophysics Research Institute, Liverpool John Moores University, Liverpool, L3 5RF, UK \\
$^{12}$ Instituto de Astrofísica, Facultad de Ciencias Exactas, Universidad Andres Bello, Fernández Concha 700, Las Condes, Santiago, Chile \\
$^{13}$ Vatican Observatory, Vatican City State 00120, Italy \\
$^{14}$ Universit\'e C\^ote d'Azur, Observatoire de la C\^ote d'Azur, CNRS, Laboratoire Lagrange, Nice, France \\
$^{15}$  Departamento de Astronomia, Casilla 160-C, Universidad de Concepcion, Chile \\
$^{16}$  Instituto de Investigaci\'on Multidisciplinario en Ciencia y Tecnología, Universidad de La Serena. Avenida Ra\'ul Bitr\'an S/N, La Serena, Chile \\
$^{17}$  Departamento de Astronomía, Facultad de Ciencias, Universidad de La Serena. Av. Juan Cisternas 1200, La Serena, Chile \\
$^{18}$ Department of Astronomy, University of Washington, Box 351580, Seattle, WA 98195, USA \\
$^{9}$ NSF’s NOIRLab, 950 N. Cherry Ave., Tucson, AZ 85719, USA \\
$^{19}$ Instituto de Astrof\'{\i}sica, Pontificia Universidad Cat\'olica de Chile, Casilla 306, Santiago 22, Chile\\
$^{20}$ Millenium Institute of Astrophysics (MAS), Nuncio Monse\~nor S\'otero Sanz 100, Providencia, Santiago Chile\\
}

\date{Accepted XXX. Received YYY; in original form ZZZ}

\pubyear{2022}

\begin{document}
\label{firstpage}
\maketitle

\begin{abstract}
The central part of the Galaxy host a multitude of stellar populations, including the spheroidal bulge stars, stars moved to the bulge through secular evolution of the bar, inner halo, inner thick disk,
inner thin disk, as well as debris from past accretion events. We identified a sample of 58 candidate stars belonging to the stellar population of the spheroidal bulge, and analyse their abundances.
The present calculations of  Mg, Ca, and Si lines are in agreement with the APOGEE-ASPCAP abundances, 
whereas abundances of C, N, O, and Ce are re-examined. 
We find normal $\alpha$-element enhancements in oxygen, similar to magnesium, 
Si, and Ca  abundances, which are typical of other bulge stars surveyed in the optical in Baade's Window. The enhancement of [O/Fe] in these stars suggests that they do not belong to accreted debris. No spread in N abundances is found, and none of the sample stars is  N-rich, indicating that these stars are not second generation stars originated in globular clusters. Ce instead is enhanced in the sample stars, which points to an s-process origin such as due to enrichment from early generations of massive fast rotating stars, the so-called spinstars. 
\end{abstract}



\section{Introduction}

The stellar populations in the central part of the Galaxy can inform us about its complex formation processes. This region was recently confirmed to  contain stars in a metal-poor spheroidal bulge (e.g. Babusiaux et al. 2010, D\'ek\'any et al. 2013, Babusiaux 2016, Zoccali et al. 2018, Savino et al. 2020, Kunder et al. 2020, Arentsen et al. 2020, Queiroz et al. 2021 and references therein),
along with a metal-rich contribution from the bar
and inner thin disk, thick disk and halo interlopers. In addition, debris of past accretion events, such as {\it{Gaia-Enceladus-Sausage}}  (GES) (Belokurov et al. 2018, Helmi et al. 2018),  and many other dwarf galaxy remnants, and minor substructures, absorbed during the early stages of the Galaxy formation (see e.g. Fern\'andez-Trincado et al. 2022, Horta et al. 2020, 2021, 2022) are present. 
Therefore studies of the Galactic bulge region are important for understanding the early stages of our Galaxy's formation (e.g., Barbuy et al. 2018a, Rojas-Arriagada et al. 2020).
In particular, Queiroz et al. (2020, 2021) combining distance derivation with proper motions from the  Gaia Early Data Release 3 (Gaia Collaboration Brown et al. 2021)  revealed stars of large
eccentricity, but with orbits confined to the bulge region -- with a maximum height from the Galactic mid-plane, |z|$_{\rm max}$, below 3 kpc, with
intermediate metallicities, which are good candidates for belonging to the
oldest Galactic bulge component (which we here call spheroid bulge stars).

The spheroidal metal-poor bulge can be thought of as a pressure supported structure formed through
violent processes, such as hierarchical clustering via minor mergers, at a very early stage of the Galaxy. Ferraro et al. (2021) finds evidence that clumps of stars and gas existed at the time of the Milky Way formation.
N-body simulations assume instead that early stellar discs heat rapidly as they form, and can lead to different density distributions for metal-rich and metal-poor stars (e.g. Debattista et al. 2017). Many other options are possible to form the metal-poor spheroid such as a major merger, accretion of dwarf galaxies, among others (e.g. Barbuy et al. 2018a). Whatever process, it leads to an observed metal-poor spheroid, and it has also
to explain the very old ages of the in-situ globular 
clusters such as e.g. HP~1 (Kerber et al. 2019), Djorgovski~2 (Ortolani
et al. 2019), Palomar~6 (Souza et al. 2021), of ages derived to be of 
12.8$\pm$0.9, 12.7$\pm$0.7, and 12.4$\pm$0.9 Gyr, respectively.

The search for the earliest stars in the Galaxy is an important endeavour to try to identify the earliest
chemical abundances imprinted in the oldest stars, and the nature of the supernovae that enriched them.
Most of the current observational efforts in finding the chemical imprints left by the first stars have focused on the most metal-poor stars found in the Milky Way halo (Beers \& Christlieb 2005; Beers et al. 2017). Very metal-poor stars were also found in ultra-faint dwarf galaxies, which are intriguing dark-matter dominated objects with very low average metallicities
(Ji et al. 2016). The Galactic bulge, as well as the halo, is a potential host
of some of the oldest stars in our Galaxy. Tumlinson (2010) suggests that half of the oldest stars were formed in the central parts of the Galaxy.
Searches for field metal-poor stars in the Galactic bulge are the target of
surveys such as those by Howes et al. (2016), Casey \& Schlaufman (2015),
the Pristine Inner Galaxy Survey (PIGS, Arentsen et al. 2020), 
HERBS (Duong et al. 2019a,b), and COMBS (Lucey et al. 2019, 2021, 2022) surveys.
Metal-poor stars in the Galactic bulge have been mostly traced by Globular clusters (Rossi et al. 2015, Bica et al. 2016), and RR Lyrae stars (Minniti et al. 2017), which show a peak at [Fe/H]$\sim-1.0$ (Barbuy et al. 2018a). 
This metallicity peak at [Fe/H]$\sim-1.0$ has been also recently confirmed regarding field stars by Lucey et al. (2021). 
In fact, it is expected that a fast chemical enrichment in the Galactic bulge results in a very old population with this relatively high metallicity, that would correspond to the age of stars with [Fe/H]$\sim-$3.0 in the halo (Chiappini et al. 2011, Wise et al. 2012, Barbuy et al. 2018a).%

Our main interest in the present work is to analyse the
abundances of stars of the spheroidal bulge with a moderate metallicity of 
[Fe/H]$< -0.8$, in order to try to identify the earliest supernovae of the central regions of the Galaxy,
and imposing constraints on the early chemical enrichment of the Milky Way.
For the selection of sample stars we applied kinematical and dynamical criteria, by combining data from APOGEE and Gaia Early Release EDR3.
We chose stars with azimuthal velocity $V_\phi < 0$ (this selection will avoid contamination by disk stars, but would still include accreted debris of objects such as GES)
that have orbits confined within 4 kpc of the Galactic center,  a maximum height of |z|$_{max} < 3.0$ kpc, eccentricity $> 0.7$, and with orbits not supporting the bar structure.
With this selection, as noted above, we expect our sample to be dominated by a pressure supported, most
probably old component of the bulge. We hope to discard the contamination of our sample by accreted debris thanks to the detailed chemical information and, in particular, the alpha-over-iron enhancement, expected to be low in most of the accreted debris.
Finally, given that we used a barred potential, the
z-component of angular momentum (Lz) 
is not conserved, and most orbits are either retrograde or prograde, and a fraction among those identified as counter-rotating keep retrograde along its orbit.

In this paper we carried out an analysis of atomic and molecular lines 
for the selected sample of 58 metal-poor spheroid bulge star candidates aiming at
refining the APOGEE Stellar Parameter and Chemical Abundance Pipeline (ASPCAP; Garc\'{\i}a-P\'erez et al. 2016) results, 
in order to interpret the derived abundances in terms of the early chemo-dynamical evolution of the bulge. As it will be shown, this re-analysis is critical 
for some alpha elements, and therefore for the identification and confirmation of  old spheroid bulge stars at moderately low metallicities.
In the present work we adopt the stellar parameters issued from the DR17
release of the APOGEE ASPCAP code. The C, N, and O abundances are derived from CO, OH and CN lines, that are interdependent, and since there are such molecular lines all over the spectra, they can affect the abundances of atomic lines. We also refine the abundances of Ce. Other elements including Na, Al and iron-peak elements will be the topic of a future work.

In Section 2 the selection of our sample is described. The element abundances are derived in Sect. 3. In Sect. 4 the results are compared with literature data for bulge samples and chemodynamical models, and discussed.
In Sect. 5 conclusions are drawn.

\section{The sample}

The Apache Point Observatory Galactic Evolution Experiment (APOGEE; Majewski et al. 2017) is part of the Sloan Digital Sky Survey 
III and IV (SDSS; Blanton et al. 2017). It is a project encompassing spectroscopic programs 
that observe Milky Way stars at high resolution and high signal-to-noise ratios (S/N) in the near-infrared (NIR).
The project SDSS-IV technical summary, the SDSS telescope and APOGEE spectrograph are  described in
Blanton et al. (2017), Gunn et al. (2006), and
 Wilson et al. (2019), respectively, whereas  
Zasowski et al. (2013, 2017), Beaton et al. (2020) 
and Santana et al. (2021) describe the  APOGEE 
and          APOGEE-2 Target Selections.
The data release 17 (DR17) contains high-resolution (R$\sim22,500$)  NIR spectra (15140-16940\,{\rm \AA}) for some 7$\times$$10^5$ stars, covering both the northern and southern sky. While APOGEE-1 observed the Milky Way bulge/bar at $l>0 \deg$, APOGEE-2 covers the whole bulge/bar region.

Given that the central part of the Milky Way hosts members of all Galactic components, including the bulge, disc, and halo (P\'erez-Villegas et al. 2020; Rojas-Arriagada et al. 2020; Queiroz et al. 2021), we have used the chemo-orbital analysis shown in Queiroz et al. (2021) to identify good candidates in the spheroidal bulge APOGEE sample. To disentangle the different stellar populations coexisting in the innermost parts of the Galaxy is not an easy task, and one of the difficulties is to compute precise distances for these stars due to the high extinction. Thanks to StarHorse (Santiago et al. 2016; Queiroz et al. 2018), precise stellar distances for the entire APOGEE sample were derived both for DR16 
(Ahumada et al. 2020, Queiroz et al. 2020), and DR17\footnote{Value added catalogues are available in both releases} (Abdurro'uf et al. 2022, Queiroz et al. 2022, in prep.).

\begin{figure}
	\includegraphics[width=\columnwidth]{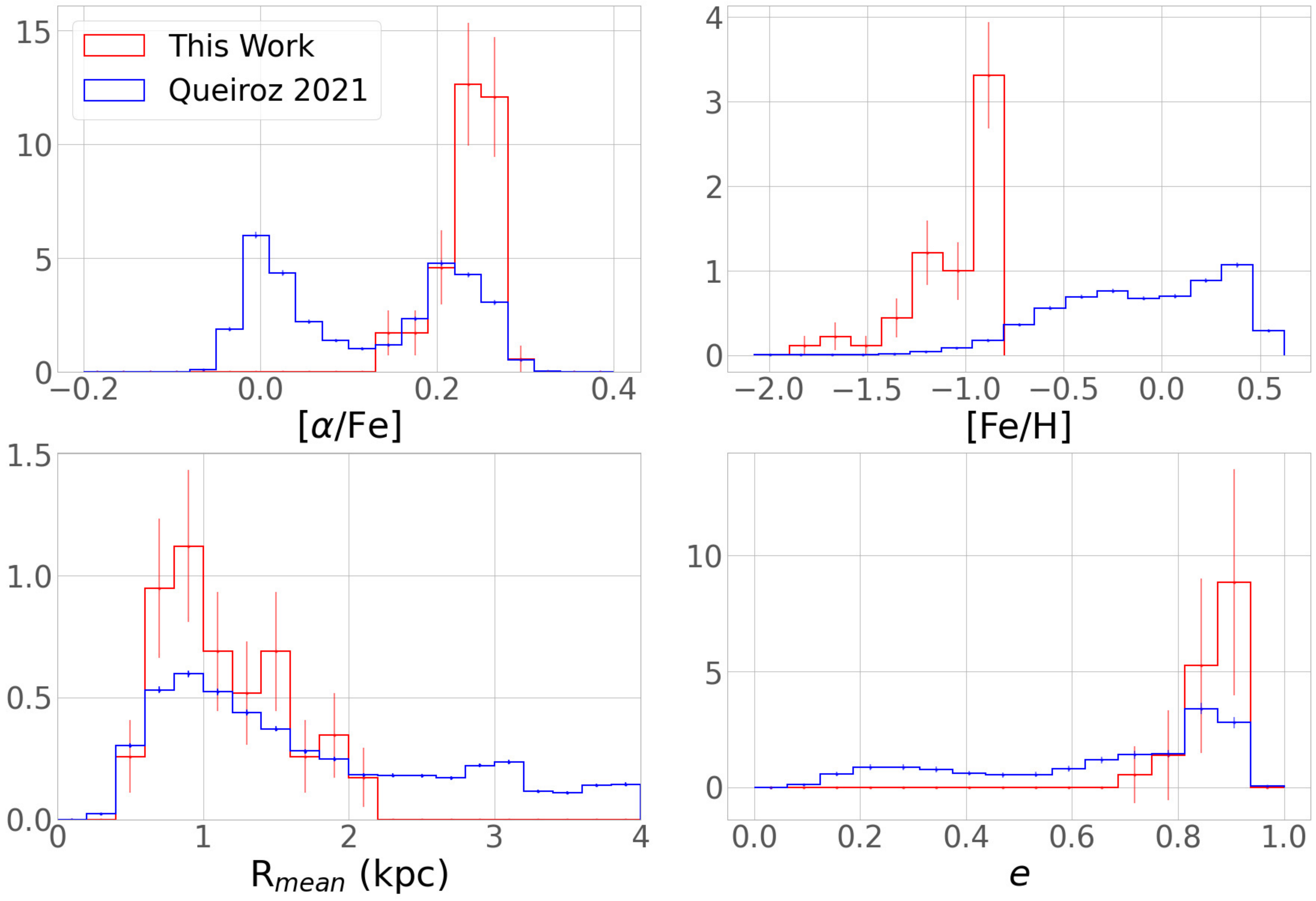}
        \caption{Comparison of the present sample of 58 selected stars (red)
         and the RPM sample of Queiroz et al. (2021) (blue). Upper panels: normalized distribution of metallicity and alpha-to-iron ratios from APOGEE; lower panels: mean radius
R$_{mean}$=((R$_{\rm apocenter}$+R$_{\rm pericenter}$)/2 and eccentricity of the orbits.}
    \label{queiroz}
\end{figure}

\begin{figure}
	\includegraphics[width=\columnwidth]{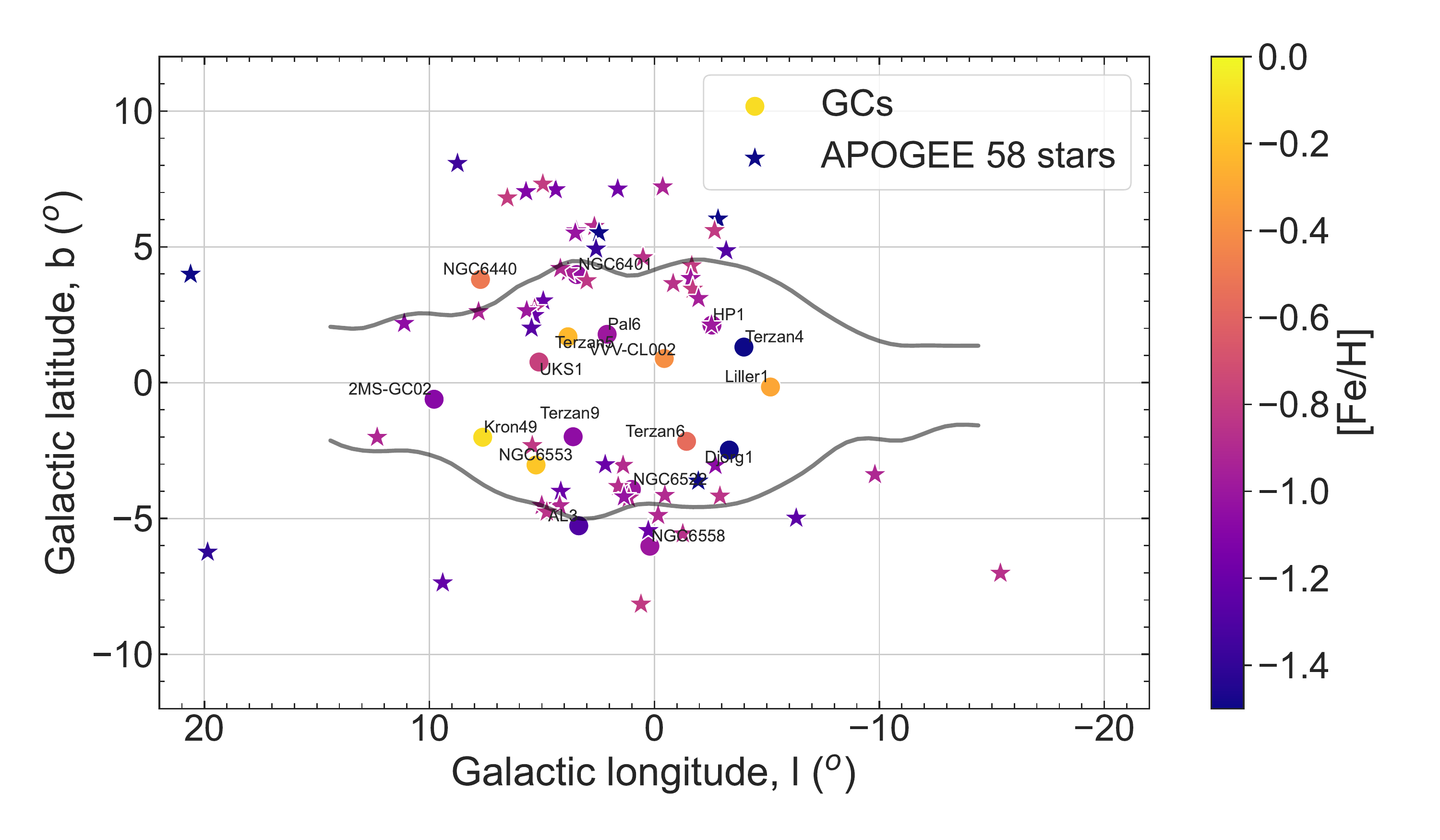}
        \caption{Projected l,b distribution of studied stars in the Galactic bulge region. Symbols: filled stars: this work; filled circles: bulge globular clusters (GCs); solid black line: contours of the bulge. The colours indicate metallicity according to the colour-bar.
        }
    \label{spatial_distribution}
\end{figure}

We selected stars from the reduced-proper-motion (RPM) sample of Queiroz et al. (2021). For that sample,
orbits were calculated using the StarHorse distances and the proper motions from the Gaia Early Data Release 3 (EDR3) (Gaia Collaboration 2021). In order to select the best candidate objects that belong to the spheroidal bulge, the following selection criteria were adopted: a maximum distance to the Galactic center of d$_{\rm GC} < 4$ kpc (Bica et al. 2016); a maximum vertical excursion from the Galactic plane $|z|_{\rm max} < 3.0$ kpc; eccentricity $> 0.7$; orbits that do not support the bar structure\footnote{To estimate this probability, we used the Monte Carlo sample of each star (50 orbits) and calculated the fraction of orbits classified as bar-shaped} (orbits with frequency ratio $f_{R}/f_{x} \neq 2.0 \pm 0.1$; Portail et al. 2015); and based on Figure 17 of Queiroz et al. (2021), we  selected counter-rotating stars ($V_{\phi} < 0.0$). Finally, according to the discussion of Sect. 1,  we considered only stars with moderate metallicity of [Fe/H] $< -0.80$. Applying the selection criteria described above, a sample of 58 stars has been selected. The adopted input parameters for the orbits integration and the orbital parameters are given in Table \ref{orbital}.
In Figure \ref{queiroz} we show the distribution of parameters for our selected stars in comparison with 
the RPM sample of Queiroz et al. (2021) and our selection is then similar to the metal-poor/high eccentricity stars
discussed in  their Fig. 20. This figure indicates that our selection is indeed reaching bulge stars of the metal-poor spheroid, that are moderately metal-poor, $\alpha$-rich  and in eccentrical orbits but confined to the Galactic center region.
Figure \ref{spatial_distribution} shows the projected l,b distribution of the
sample in the Galactic bulge region.

As explained above, our stars were selected from the reduced-proper-motion sample of Queiroz et al. (2021), and therefore have a signal-to-noise SNR$>$50, a good spectral fit from the ASPCAP pipeline ASPCAP$_{-}$Chi2$<$ 25,  and a radial velocity scatter Vscatter$<$1.5 km s$^{-1}$. 
As for the renormalized unit weight error - RUWE Gaia EDR3 parameter, 56 out of 58 stars in our sample comply with the standard or minimal requirements to get reliable orbital elements, since astrometry from Gaia EDR3 has its own caveats. According to the Gaia consortium, the RUWE parameter is suggested to return stars astrometrically well-behaved by applying a cut with 
 RUWE$\leq$1.4, which is followed by the 56 stars listed in Table \ref{orbital}.
The stars 2M17453659-2309130 and 2M18023156-2834451 have a RUWE $>$ 1.4, which makes them sources with astrometric parameters that are not reliable enough.

\begin{figure}
	\includegraphics[width=\columnwidth]{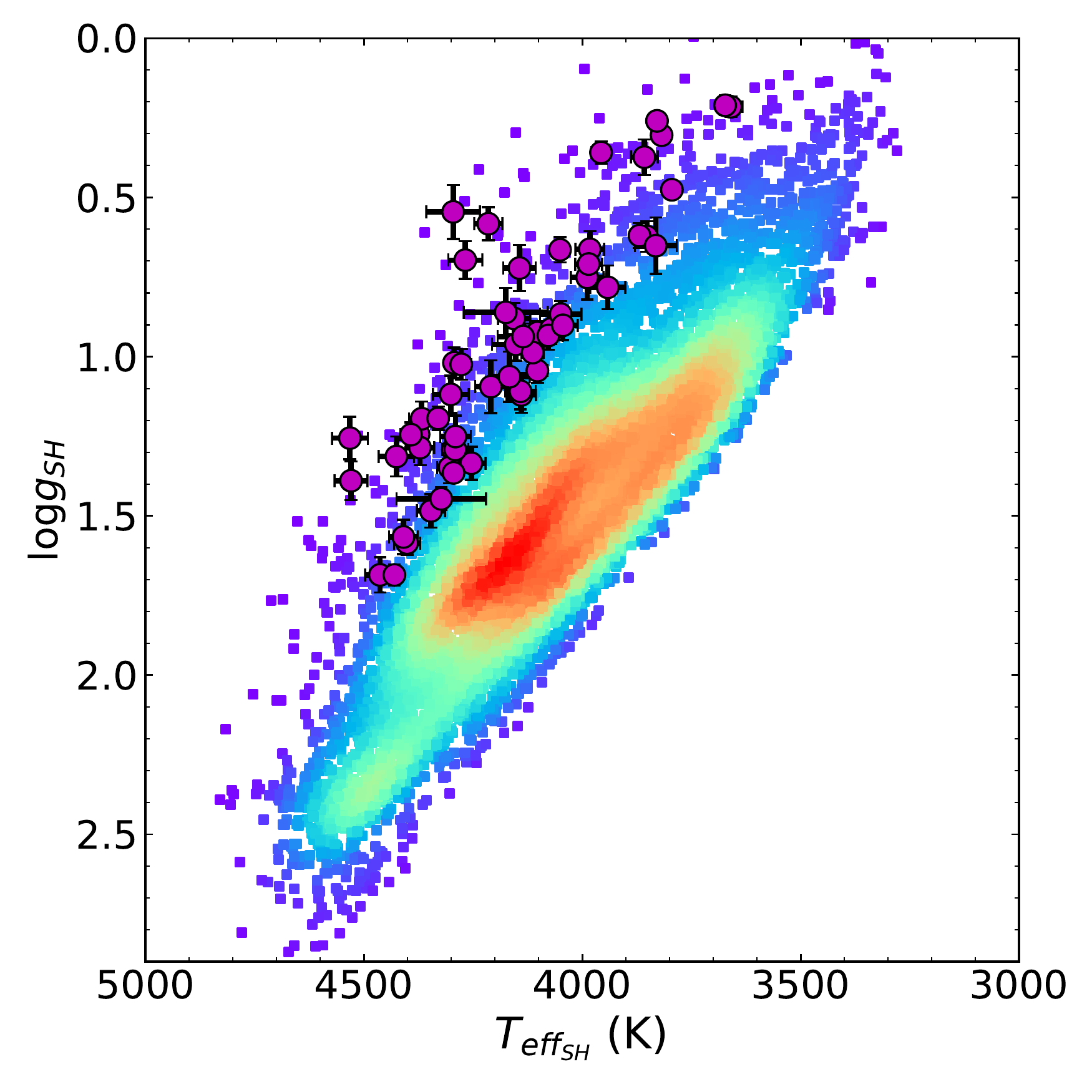}
    \caption{Kiel diagram of the 58 sample bulge stars.
    (purple circles). In the background, we show the full reduced proper-motion sample
    of Queiroz et al. (2021).}
    \label{cmd}
\end{figure}

In Figure \ref{cmd}  a Kiel diagram of the sample stars is plotted with the  effective temperature from Apogee-ASPCAP and gravity log g coming from the StarHorse output from Queiroz et al. (2020), and compared with the reduced-proper-motion sample of Queiroz et al. (2021).

\begin{table*}
\caption[4]{Coordinates, Starshorse distances, Gaia EDR3 proper motions, Gaia DR2 radial velocity, and orbital parameters for the selected 58 stars from RPM sample of Queiroz et al. (2021). }
\label{orbital}
\begin{tabular}{l@{\hspace{6pt}}c@{\hspace{5pt}}c@{\hspace{6pt}}c@{\hspace{5pt}}c@{\hspace{5pt}}c@{\hspace{5pt}}c@{\hspace{5pt}}c@{\hspace{5pt}}c@{\hspace{6pt}}c@{\hspace{6pt}}c@{\hspace{6pt}}}
\noalign{\smallskip}
\hline
\noalign{\smallskip}
\hbox{ID}& 
$\alpha$ & $\delta$ & d$_\odot$ &  $\mu_{\alpha}^*$ & $\mu_{\delta}$ & RV & $r_{\rm min}$ &  $r_{\rm max}$ 
& $|z|_{\rm max}$ & $e$\\
 & \hbox{(º)} & \hbox{(º)} & \hbox{(kpc)} &  \hbox{(mas yr$^{-1}$)} & \hbox{(mas yr$^{-1}$)} & (km s$^{-1}$)& (kpc)&(kpc) &(kpc) &  \\  

 \noalign{\smallskip}
\noalign{\hrule}
\hline

 2M17153858-2759467 & $ 258.911$ & $ -27.996$ & $    8.51 \pm     0.50$ & $   -5.46 \pm     0.02$ & $   -5.30 \pm     0.02$ & $  191.79 \pm     0.01$ & $    0.13 \pm     0.05$ & $    2.51 \pm     0.46$ & $    1.68 \pm     0.18$ & $    0.90 \pm     0.04$ \\ 
  2M17173248-2518529 & $ 259.385$ & $ -25.315$ & $    7.79 \pm     0.91$ & $   -2.14 \pm     0.04$ & $   -9.47 \pm     0.03$ & $  187.54 \pm     0.02$ & $    0.19 \pm     0.16$ & $    3.73 \pm     0.79$ & $    2.34 \pm     0.49$ & $    0.91 \pm     0.05$ \\ 
  2M17173693-2806495 & $ 259.404$ & $ -28.114$ & $    6.94 \pm     0.45$ & $   -4.85 \pm     0.03$ & $   -9.80 \pm     0.02$ & $ -104.63 \pm     0.01$ & $    0.12 \pm     0.06$ & $    1.94 \pm     0.44$ & $    1.52 \pm     0.19$ & $    0.89 \pm     0.05$ \\ 
  2M17190320-2857321 & $ 259.763$ & $ -28.959$ & $    6.81 \pm     0.46$ & $   -5.95 \pm     0.03$ & $   -7.60 \pm     0.02$ & $  -83.87 \pm     0.03$ & $    0.14 \pm     0.07$ & $    1.80 \pm     0.33$ & $    0.74 \pm     0.26$ & $    0.87 \pm     0.05$ \\ 
  2M17224443-2343053 & $ 260.685$ & $ -23.718$ & $    6.02 \pm     0.42$ & $   -9.20 \pm     0.02$ & $   -8.15 \pm     0.01$ & $  114.23 \pm     0.01$ & $    0.21 \pm     0.12$ & $    3.87 \pm     0.51$ & $    2.61 \pm     0.23$ & $    0.89 \pm     0.07$ \\ 
  2M17250290-2800385 & $ 261.262$ & $ -28.011$ & $    5.83 \pm     0.76$ & $   -3.05 \pm     0.03$ & $   -9.26 \pm     0.02$ & $   26.27 \pm     0.01$ & $    0.20 \pm     0.11$ & $    2.63 \pm     0.71$ & $    1.02 \pm     0.38$ & $    0.85 \pm     0.07$ \\ 
  2M17265563-2813558 & $ 261.732$ & $ -28.232$ & $    7.55 \pm     0.56$ & $   -7.25 \pm     0.04$ & $   -7.31 \pm     0.03$ & $  196.52 \pm     0.03$ & $    0.13 \pm     0.07$ & $    2.40 \pm     0.53$ & $    1.52 \pm     0.36$ & $    0.91 \pm     0.04$ \\ 
  2M17281191-2831393 & $ 262.050$ & $ -28.528$ & $    6.50 \pm     0.58$ & $   -9.70 \pm     0.03$ & $   -4.61 \pm     0.02$ & $   81.01 \pm     0.02$ & $    0.14 \pm     0.05$ & $    2.24 \pm     0.68$ & $    1.87 \pm     0.29$ & $    0.90 \pm     0.05$ \\ 
  2M17285088-2855427 & $ 262.212$ & $ -28.929$ & $    7.59 \pm     0.42$ & $   -4.80 \pm     0.03$ & $   -5.57 \pm     0.02$ & $   -7.43 \pm     0.01$ & $    0.06 \pm     0.03$ & $    0.83 \pm     0.25$ & $    0.47 \pm     0.02$ & $    0.87 \pm     0.04$ \\ 
  2M17291778-2602468 & $ 262.324$ & $ -26.046$ & $    6.93 \pm     0.44$ & $   -5.60 \pm     0.06$ & $   -7.06 \pm     0.04$ & $  -47.65 \pm     0.01$ & $    0.12 \pm     0.07$ & $    1.50 \pm     0.43$ & $    0.66 \pm     0.10$ & $    0.86 \pm     0.07$ \\ 
  2M17292082-2126433 & $ 262.337$ & $ -21.445$ & $    6.60 \pm     0.68$ & $   -0.84 \pm     0.02$ & $  -10.79 \pm     0.02$ & $  -79.08 \pm     0.01$ & $    0.19 \pm     0.10$ & $    2.83 \pm     0.67$ & $    2.06 \pm     0.14$ & $    0.87 \pm     0.09$ \\ 
  2M17293482-2741164 & $ 262.395$ & $ -27.688$ & $    6.81 \pm     0.52$ & $   -3.56 \pm     0.04$ & $   -8.16 \pm     0.03$ & $  -74.26 \pm     0.02$ & $    0.09 \pm     0.06$ & $    1.51 \pm     0.39$ & $    0.59 \pm     0.10$ & $    0.89 \pm     0.05$ \\ 
  2M17295481-2051262 & $ 262.478$ & $ -20.857$ & $    7.00 \pm     0.38$ & $    0.11 \pm     0.04$ & $   -6.20 \pm     0.03$ & $ -213.15 \pm     0.04$ & $    0.16 \pm     0.08$ & $    3.43 \pm     0.40$ & $    2.31 \pm     0.08$ & $    0.90 \pm     0.04$ \\ 
  2M17301495-2337002 & $ 262.562$ & $ -23.617$ & $    8.28 \pm     0.66$ & $   -8.24 \pm     0.04$ & $   -9.11 \pm     0.02$ & $  -70.19 \pm     0.01$ & $    0.24 \pm     0.17$ & $    1.97 \pm     0.93$ & $    1.81 \pm     0.32$ & $    0.83 \pm     0.17$ \\ 
  2M17303581-2354453 & $ 262.649$ & $ -23.913$ & $    7.99 \pm     0.60$ & $   -8.31 \pm     0.04$ & $   -4.45 \pm     0.02$ & $   27.88 \pm     0.01$ & $    0.10 \pm     0.14$ & $    1.52 \pm     0.40$ & $    1.38 \pm     0.18$ & $    0.87 \pm     0.08$ \\ 
  2M17310874-2956542 & $ 262.786$ & $ -29.948$ & $    6.81 \pm     0.00$ & $   -3.38 \pm     0.04$ & $   -7.93 \pm     0.03$ & $  -10.11 \pm     0.02$ & $    0.19 \pm     0.01$ & $    1.50 \pm     0.01$ & $    0.36 \pm     0.00$ & $    0.77 \pm     0.01$ \\ 
  2M17323787-2023013 & $ 263.158$ & $ -20.384$ & $    7.76 \pm     0.55$ & $   -5.22 \pm     0.03$ & $   -1.34 \pm     0.02$ & $  -97.24 \pm     0.01$ & $    0.17 \pm     0.08$ & $    2.53 \pm     0.24$ & $    1.62 \pm     0.22$ & $    0.88 \pm     0.06$ \\ 
  2M17324257-2301417 & $ 263.177$ & $ -23.028$ & $    7.69 \pm     0.74$ & $   -2.70 \pm     0.05$ & $   -7.92 \pm     0.03$ & $ -181.81 \pm     0.01$ & $    0.16 \pm     0.13$ & $    1.67 \pm     0.58$ & $    1.27 \pm     0.28$ & $    0.83 \pm     0.09$ \\ 
  2M17330695-2302130 & $ 263.279$ & $ -23.037$ & $    7.40 \pm     0.10$ & $   -3.51 \pm     0.04$ & $   -9.38 \pm     0.03$ & $    6.42 \pm     0.00$ & $    0.11 \pm     0.05$ & $    1.44 \pm     0.04$ & $    0.95 \pm     0.01$ & $    0.86 \pm     0.05$ \\ 
  2M17330730-2407378 & $ 263.280$ & $ -24.127$ & $    5.32 \pm     0.25$ & $   -4.74 \pm     0.03$ & $   -8.85 \pm     0.02$ & $  -31.23 \pm     0.01$ & $    0.11 \pm     0.04$ & $    3.15 \pm     0.25$ & $    1.41 \pm     0.42$ & $    0.93 \pm     0.03$ \\ 
  2M17341796-3905103 & $ 263.575$ & $ -39.086$ & $    8.63 \pm     0.69$ & $   -2.19 \pm     0.07$ & $   -3.37 \pm     0.05$ & $    3.77 \pm     0.03$ & $    0.23 \pm     0.06$ & $    1.90 \pm     0.22$ & $    0.65 \pm     0.27$ & $    0.78 \pm     0.05$ \\ 
  2M17342067-3902066 & $ 263.586$ & $ -39.035$ & $    9.80 \pm     0.00$ & $   -2.51 \pm     0.08$ & $   -3.17 \pm     0.06$ & $    5.95 \pm     0.04$ & $    0.13 \pm     0.05$ & $    2.50 \pm     0.14$ & $    1.50 \pm     0.18$ & $    0.90 \pm     0.04$ \\ 
  2M17344841-4540171 & $ 263.702$ & $ -45.671$ & $    6.71 \pm     0.38$ & $   -0.85 \pm     0.02$ & $   -6.51 \pm     0.01$ & $  148.00 \pm     0.01$ & $    0.17 \pm     0.15$ & $    3.67 \pm     0.51$ & $    2.61 \pm     0.18$ & $    0.91 \pm     0.09$ \\ 
  2M17351981-1948329 & $ 263.833$ & $ -19.809$ & $    8.20 \pm     0.32$ & $   -2.39 \pm     0.02$ & $   -6.57 \pm     0.01$ & $ -230.13 \pm     0.00$ & $    0.36 \pm     0.23$ & $    2.61 \pm     0.53$ & $    2.14 \pm     0.16$ & $    0.77 \pm     0.16$ \\ 
  2M17354093-1716200 & $ 263.921$ & $ -17.272$ & $    6.15 \pm     0.35$ & $   -4.18 \pm     0.02$ & $   -7.53 \pm     0.01$ & $  -84.29 \pm     0.01$ & $    0.17 \pm     0.11$ & $    2.84 \pm     0.27$ & $    1.59 \pm     0.16$ & $    0.88 \pm     0.06$ \\ 
  2M17382504-2424163 & $ 264.604$ & $ -24.405$ & $    6.78 \pm     0.52$ & $   -2.34 \pm     0.07$ & $   -8.58 \pm     0.04$ & $  -56.51 \pm     0.01$ & $    0.12 \pm     0.09$ & $    1.68 \pm     0.43$ & $    0.66 \pm     0.08$ & $    0.87 \pm     0.10$ \\ 
  2M17390801-2331379 & $ 264.783$ & $ -23.527$ & $    7.57 \pm     0.54$ & $   -7.05 \pm     0.03$ & $   -3.91 \pm     0.02$ & $ -199.67 \pm     0.01$ & $    0.13 \pm     0.07$ & $    1.92 \pm     0.36$ & $    1.43 \pm     0.17$ & $    0.88 \pm     0.05$ \\ 
  2M17392719-2310311 & $ 264.863$ & $ -23.175$ & $    6.70 \pm     0.31$ & $  -10.26 \pm     0.03$ & $   -7.39 \pm     0.02$ & $   47.66 \pm     0.00$ & $    0.13 \pm     0.07$ & $    2.67 \pm     0.47$ & $    1.90 \pm     0.22$ & $    0.90 \pm     0.04$ \\ 
  2M17453659-2309130 & $ 266.402$ & $ -23.154$ & $    6.31 \pm     0.56$ & $   -4.98 \pm     0.23$ & $   -7.39 \pm     0.15$ & $ -140.43 \pm     0.02$ & $    0.12 \pm     0.04$ & $    2.21 \pm     0.47$ & $    0.54 \pm     0.32$ & $    0.90 \pm     0.05$ \\ 
  2M17473299-2258254 & $ 266.887$ & $ -22.974$ & $    7.36 \pm     0.61$ & $   -4.18 \pm     0.02$ & $   -9.24 \pm     0.01$ & $  -39.26 \pm     0.01$ & $    0.11 \pm     0.04$ & $    1.37 \pm     0.35$ & $    0.44 \pm     0.04$ & $    0.87 \pm     0.05$ \\ 
  2M17482995-2305299 & $ 267.125$ & $ -23.092$ & $    7.05 \pm     0.43$ & $   -0.95 \pm     0.03$ & $   -6.72 \pm     0.02$ & $ -216.54 \pm     0.02$ & $    0.14 \pm     0.06$ & $    2.07 \pm     0.51$ & $    0.73 \pm     0.32$ & $    0.87 \pm     0.06$ \\ 
  2M17483633-2242483 & $ 267.151$ & $ -22.713$ & $    8.11 \pm     0.69$ & $   -0.62 \pm     0.03$ & $   -9.74 \pm     0.02$ & $  -93.04 \pm     0.00$ & $    0.12 \pm     0.07$ & $    1.38 \pm     0.61$ & $    0.87 \pm     0.34$ & $    0.85 \pm     0.09$ \\ 
  2M17503065-2313234 & $ 267.628$ & $ -23.223$ & $    6.83 \pm     0.38$ & $   -4.88 \pm     0.05$ & $   -6.57 \pm     0.03$ & $ -203.16 \pm     0.01$ & $    0.09 \pm     0.03$ & $    1.94 \pm     0.42$ & $    0.37 \pm     0.13$ & $    0.92 \pm     0.03$ \\ 
  2M17503263-3654102 & $ 267.636$ & $ -36.903$ & $    7.49 \pm     0.62$ & $   -7.00 \pm     0.02$ & $   -4.97 \pm     0.01$ & $   11.58 \pm     0.01$ & $    0.10 \pm     0.05$ & $    1.51 \pm     0.38$ & $    1.28 \pm     0.08$ & $    0.89 \pm     0.04$ \\ 
  2M17511568-3249403 & $ 267.815$ & $ -32.828$ & $    7.54 \pm     0.59$ & $   -4.58 \pm     0.04$ & $   -9.25 \pm     0.03$ & $ -102.21 \pm     0.01$ & $    0.08 \pm     0.03$ & $    1.28 \pm     0.21$ & $    0.45 \pm     0.03$ & $    0.88 \pm     0.04$ \\ 
  2M17532599-2053304 & $ 268.358$ & $ -20.892$ & $    7.66 \pm     0.59$ & $   -3.44 \pm     0.04$ & $   -7.77 \pm     0.03$ & $  -78.10 \pm     0.01$ & $    0.08 \pm     0.04$ & $    1.39 \pm     0.22$ & $    0.42 \pm     0.05$ & $    0.89 \pm     0.04$ \\ 
  2M17552681-3334272 & $ 268.862$ & $ -33.574$ & $    7.67 \pm     0.55$ & $   -3.57 \pm     0.03$ & $   -4.88 \pm     0.02$ & $  166.48 \pm     0.02$ & $    0.09 \pm     0.03$ & $    1.43 \pm     0.25$ & $    0.65 \pm     0.04$ & $    0.89 \pm     0.04$ \\ 
  2M17552744-3228019 & $ 268.864$ & $ -32.467$ & $    7.10 \pm     0.89$ & $   -7.00 \pm     0.03$ & $   -6.81 \pm     0.02$ & $  -71.82 \pm     0.01$ & $    0.10 \pm     0.05$ & $    1.34 \pm     0.68$ & $    0.81 \pm     0.24$ & $    0.88 \pm     0.06$ \\ 
  2M18005152-2916576 & $ 270.215$ & $ -29.283$ & $    8.45 \pm     0.60$ & $    1.18 \pm     0.04$ & $   -9.34 \pm     0.03$ & $  -77.43 \pm     0.02$ & $    0.17 \pm     0.09$ & $    1.25 \pm     0.61$ & $    1.07 \pm     0.34$ & $    0.80 \pm     0.10$ \\ 
  2M18010424-3126158 & $ 270.268$ & $ -31.438$ & $    7.10 \pm     0.57$ & $   -1.22 \pm     0.03$ & $   -9.10 \pm     0.02$ & $   81.96 \pm     0.00$ & $    0.09 \pm     0.04$ & $    1.43 \pm     0.50$ & $    0.82 \pm     0.15$ & $    0.88 \pm     0.06$ \\ 
  2M18020063-1814495 & $ 270.503$ & $ -18.247$ & $    5.97 \pm     0.38$ & $   -4.65 \pm     0.05$ & $   -8.19 \pm     0.04$ & $  -94.07 \pm     0.02$ & $    0.11 \pm     0.03$ & $    2.85 \pm     0.40$ & $    0.55 \pm     0.44$ & $    0.92 \pm     0.02$ \\ 
  2M18023156-2834451 & $ 270.632$ & $ -28.579$ & $    8.15 \pm     0.44$ & $   -4.55 \pm     0.07$ & $  -10.32 \pm     0.05$ & $ -190.17 \pm     0.01$ & $    0.27 \pm     0.10$ & $    1.59 \pm     0.40$ & $    0.62 \pm     0.33$ & $    0.73 \pm     0.12$ \\ 
  2M18042687-2928348 & $ 271.112$ & $ -29.476$ & $    7.89 \pm     0.75$ & $   -2.34 \pm     0.03$ & $   -7.82 \pm     0.02$ & $ -113.51 \pm     0.02$ & $    0.08 \pm     0.06$ & $    1.05 \pm     0.41$ & $    0.61 \pm     0.07$ & $    0.89 \pm     0.07$ \\ 
  2M18044663-3132174 & $ 271.194$ & $ -31.538$ & $    7.31 \pm     0.43$ & $   -6.68 \pm     0.03$ & $   -7.25 \pm     0.02$ & $ -145.20 \pm     0.01$ & $    0.10 \pm     0.04$ & $    1.62 \pm     0.30$ & $    1.10 \pm     0.14$ & $    0.89 \pm     0.05$ \\ 
  2M18050452-3249149 & $ 271.269$ & $ -32.821$ & $    5.51 \pm     0.41$ & $   -3.19 \pm     0.02$ & $  -10.36 \pm     0.01$ & $   46.90 \pm     0.01$ & $    0.12 \pm     0.06$ & $    3.58 \pm     0.59$ & $    1.47 \pm     0.51$ & $    0.93 \pm     0.04$ \\ 
  2M18050663-3005419 & $ 271.278$ & $ -30.095$ & $    7.92 \pm     0.36$ & $   -1.98 \pm     0.04$ & $   -8.42 \pm     0.03$ & $ -137.47 \pm     0.00$ & $    0.09 \pm     0.04$ & $    1.09 \pm     0.11$ & $    0.77 \pm     0.05$ & $    0.86 \pm     0.07$ \\ 
  2M18052388-2953056 & $ 271.350$ & $ -29.885$ & $    7.43 \pm     0.59$ & $   -5.77 \pm     0.03$ & $   -8.14 \pm     0.02$ & $   -4.77 \pm     0.05$ & $    0.09 \pm     0.06$ & $    1.11 \pm     0.26$ & $    0.66 \pm     0.03$ & $    0.85 \pm     0.06$ \\ 
  2M18065321-2524392 & $ 271.722$ & $ -25.411$ & $    7.91 \pm     0.80$ & $   -7.64 \pm     0.06$ & $   -8.61 \pm     0.04$ & $ -112.08 \pm     0.01$ & $    0.28 \pm     0.14$ & $    1.71 \pm     0.83$ & $    0.60 \pm     0.46$ & $    0.74 \pm     0.13$ \\ 
  2M18080306-3125381 & $ 272.013$ & $ -31.427$ & $   10.06 \pm     0.73$ & $   -1.89 \pm     0.05$ & $   -4.50 \pm     0.04$ & $   23.35 \pm     0.03$ & $    0.13 \pm     0.08$ & $    2.39 \pm     0.80$ & $    1.38 \pm     0.43$ & $    0.89 \pm     0.05$ \\ 
  2M18104496-2719514 & $ 272.687$ & $ -27.331$ & $    7.30 \pm     0.31$ & $   -1.79 \pm     0.03$ & $   -7.09 \pm     0.03$ & $ -163.54 \pm     0.02$ & $    0.12 \pm     0.05$ & $    1.57 \pm     0.27$ & $    0.63 \pm     0.05$ & $    0.87 \pm     0.05$ \\ 
  2M18125718-2732215 & $ 273.238$ & $ -27.539$ & $    8.12 \pm     0.34$ & $   -5.57 \pm     0.02$ & $   -7.99 \pm     0.02$ & $  -86.39 \pm     0.00$ & $    0.13 \pm     0.04$ & $    1.14 \pm     0.16$ & $    0.79 \pm     0.09$ & $    0.80 \pm     0.07$ \\ 
  2M18142265-0904155 & $ 273.594$ & $  -9.071$ & $    6.92 \pm     0.26$ & $   -1.42 \pm     0.11$ & $   -8.72 \pm     0.09$ & $ -151.19 \pm     0.02$ & $    0.28 \pm     0.13$ & $    3.68 \pm     0.28$ & $    2.50 \pm     0.59$ & $    0.87 \pm     0.06$ \\ 
  2M18143710-2650147 & $ 273.655$ & $ -26.837$ & $    7.46 \pm     0.52$ & $   -3.66 \pm     0.04$ & $   -7.43 \pm     0.04$ & $ -200.72 \pm     0.02$ & $    0.16 \pm     0.13$ & $    1.62 \pm     0.38$ & $    1.07 \pm     0.25$ & $    0.85 \pm     0.14$ \\ 
  2M18150516-2708486 & $ 273.772$ & $ -27.147$ & $    6.80 \pm     0.38$ & $   -0.00 \pm     0.03$ & $   -9.08 \pm     0.02$ & $ -141.63 \pm     0.02$ & $    0.13 \pm     0.05$ & $    2.28 \pm     0.35$ & $    1.25 \pm     0.22$ & $    0.88 \pm     0.04$ \\ 
  2M18195859-1912513 & $ 274.994$ & $ -19.214$ & $    6.07 \pm     0.29$ & $   -6.28 \pm     0.05$ & $   -6.91 \pm     0.03$ & $  -78.88 \pm     0.02$ & $    0.10 \pm     0.05$ & $    2.81 \pm     0.24$ & $    0.76 \pm     0.37$ & $    0.93 \pm     0.03$ \\ 
  2M18200365-3224168 & $ 275.015$ & $ -32.405$ & $    6.27 \pm     0.40$ & $   -4.20 \pm     0.03$ & $  -10.36 \pm     0.02$ & $ -124.55 \pm     0.01$ & $    0.18 \pm     0.08$ & $    3.35 \pm     0.47$ & $    2.19 \pm     0.21$ & $    0.91 \pm     0.05$ \\ 
  2M18344461-2415140 & $ 278.686$ & $ -24.254$ & $    7.46 \pm     0.50$ & $   -3.96 \pm     0.03$ & $   -8.27 \pm     0.02$ & $ -171.66 \pm     0.02$ & $    0.23 \pm     0.20$ & $    2.63 \pm     0.49$ & $    2.12 \pm     0.12$ & $    0.85 \pm     0.14$ \\ 
  2M18500307-1427291 & $ 282.513$ & $ -14.458$ & $    6.40 \pm     0.32$ & $    0.13 \pm     0.03$ & $   -6.71 \pm     0.02$ & $ -134.03 \pm     0.02$ & $    0.16 \pm     0.07$ & $    3.92 \pm     0.29$ & $    2.49 \pm     0.28$ & $    0.92 \pm     0.03$ \\

\hline
\noalign{\smallskip}

\hline 
\end{tabular}
\end{table*}

\begin{table}
\caption[4]{\label{param58_new}
Selected 58 stars and corresponding DR17 non-calibrated stellar parameters.}
\begin{tabular}{lcccc|cccc}
\noalign{\smallskip}
\hline
\noalign{\smallskip}
\hbox{ID}&  T$_{\rm eff(nc)}$ &\hbox{log~g}$_{\rm (nc)}$ & \hbox{[Fe/H]}$_{\rm (nc)}$ &  \hbox{v$_t$}\\

 &  \hbox{(K)} &  &  & \hbox{(km/s)} \\  

 \noalign{\smallskip}
\noalign{\hrule}
\hline
2M17153858-2759467 & \phantom{-} 3922.7 & \phantom{-} 0.34 & \phantom{-} -1.62 & 2.62\\
  2M17173693-2806495 & \phantom{-} 3908.9 & \phantom{-} 0.95 & \phantom{-} -0.97 & 2.20\\
  2M17250290-2800385 & \phantom{-} 3796.6 & \phantom{-} 0.91 & \phantom{-} -0.80 & 2.39\\
  2M17265563-2813558 & \phantom{-} 4096.2 & \phantom{-} 1.0  & \phantom{-} -1.31 & 1.89\\
  2M17281191-2831393 & \phantom{-} 4029.1 & \phantom{-} 0.95 & \phantom{-} -1.17 & 1.73\\
  2M17295481-2051262 & \phantom{-} 4205.9 & \phantom{-} 1.50 & \phantom{-} -0.85 & 1.71\\
  2M17303581-2354453 & \phantom{-} 3863.0 & \phantom{-} 0.77 & \phantom{-} -0.99 & 2.13\\
  2M17324257-2301417 & \phantom{-} 3668.2 & \phantom{-} 0.79 & \phantom{-} -0.82 & 2.30\\
  2M17330695-2302130 & \phantom{-} 3566.6 & \phantom{-} 0.35 & \phantom{-} -0.93 & 2.42\\
  2M17344841-4540171 & \phantom{-} 3869.2 & \phantom{-} 0.85 & \phantom{-} -0.88 & 2.16\\
  2M17351981-1948329 & \phantom{-} 3553.5 & \phantom{-} 0.44 & \phantom{-} -1.11 & 3.06\\
  2M17354093-1716200 & \phantom{-} 3895.5 & \phantom{-} 1.01 & \phantom{-} -0.87 & 2.01\\
  2M17390801-2331379 & \phantom{-} 3740.4 & \phantom{-} 0.83 & \phantom{-} -0.81 & 2.34\\
  2M17392719-2310311 & \phantom{-} 3643.3 & \phantom{-} 0.67 & \phantom{-} -0.87 & 2.55\\
  2M17473299-2258254 & \phantom{-} 4018.3 & \phantom{-} 0.47 & \phantom{-} -1.71 & 2.12\\
  2M17482995-2305299 & \phantom{-} 4213.6 & \phantom{-} 1.24 & \phantom{-} -1.01 & 2.10\\
  2M17483633-2242483 & \phantom{-} 3651.5 & \phantom{-} 0.44 & \phantom{-} -1.09 & 2.57\\
  2M17503263-3654102 & \phantom{-} 3893.5 & \phantom{-} 0.64 & \phantom{-} -0.99 & 2.19\\
  2M17552744-3228019 & \phantom{-} 4018.9 & \phantom{-} 1.0  & \phantom{-} -1.05 & 1.99\\
  2M18020063-1814495 & \phantom{-} 3988.8 & \phantom{-} 0.80 & \phantom{-} -1.38 & 2.04\\
  2M18050452-3249149 & \phantom{-} 3940.8 & \phantom{-} 0.77 & \phantom{-} -1.16 & 2.08\\
  2M18050663-3005419 & \phantom{-} 3439.9 & \phantom{-} 0.23 & \phantom{-} -0.92 & 2.52\\
  2M18065321-2524392 & \phantom{-} 3893.1 & \phantom{-} 0.95 & \phantom{-} -0.89 & 2.02\\
  2M18104496-2719514 & \phantom{-} 4153.1 & \phantom{-} 1.33 & \phantom{-} -0.82 & 2.05\\
  2M18125718-2732215 & \phantom{-} 3617.2 & \phantom{-} 0.44 & \phantom{-} -1.31 & 2.64\\
  2M18200365-3224168 & \phantom{-} 3976.6 & \phantom{-} 0.95 & \phantom{-} -0.86 & 1.94\\
  2M18500307-1427291 & \phantom{-} 4076.0 & \phantom{-} 1.23 & \phantom{-} -0.94 & 1.73\\
  2M17173248-2518529 & \phantom{-} 3977.0 & \phantom{-} 1.0  & \phantom{-} -0.91 & 1.81\\
  2M17285088-2855427 & \phantom{-} 3838.0 & \phantom{-} 0.63 & \phantom{-} -1.20 & 2.18\\
  2M17291778-2602468 & \phantom{-} 3844.3 & \phantom{-} 0.71 & \phantom{-} -0.99 & 2.10\\
  2M17301495-2337002 & \phantom{-} 3814.0 & \phantom{-} 0.69 & \phantom{-} -1.06 & 2.22\\
  2M17310874-2956542 & \phantom{-} 4175.7 & \phantom{-} 1.19 & \phantom{-} -0.92 & 2.07\\
  2M17382504-2424163 & \phantom{-} 3880.4 & \phantom{-} 0.99 & \phantom{-} -1.05 & 1.55\\
  2M17453659-2309130 & \phantom{-} 4133.1 & \phantom{-} 1.27 & \phantom{-} -1.20 & 1.08\\
  2M17511568-3249403 & \phantom{-} 3921.2 & \phantom{-} 0.98 & \phantom{-} -0.90 & 2.04\\
  2M17532599-2053304 & \phantom{-} 3896.9 & \phantom{-} 0.91 & \phantom{-} -0.87 & 2.10\\
  2M17552681-3334272 & \phantom{-} 4051.0 & \phantom{-} 1.08 & \phantom{-} -0.89 & 1.98\\
  2M18005152-2916576 & \phantom{-} 4158.9 & \phantom{-} 1.04 & \phantom{-} -1.02 & 2.21\\
  2M18010424-3126158 & \phantom{-} 3773.1 & \phantom{-} 0.68 & \phantom{-} -0.83 & 2.20\\
  2M18042687-2928348 & \phantom{-} 4164.7 & \phantom{-} 0.88 & \phantom{-} -1.19 & 2.14\\
  2M18044663-3132174 & \phantom{-} 3832.6 & \phantom{-} 0.92 & \phantom{-} -0.90 & 2.22\\
  2M18052388-2953056 & \phantom{-} 4252.9 & \phantom{-} 0.92 & \phantom{-} -1.56 & 1.92\\
  2M18080306-3125381 & \phantom{-} 4310.0 & \phantom{-} 1.57 & \phantom{-} -0.90 & 1.48\\
  2M18142265-0904155 & \phantom{-} 3920.5 & \phantom{-} 1.12 & \phantom{-} -0.85 & 2.13\\
  2M18195859-1912513 & \phantom{-} 4102.0 & \phantom{-} 1.05 & \phantom{-} -1.22 & 1.78\\
  2M17190320-2857321 & \phantom{-} 4139.6 & \phantom{-} 1.19 & \phantom{-} -1.20 & 1.83\\
  2M17224443-2343053 & \phantom{-} 4058.3 & \phantom{-} 1.02 & \phantom{-} -0.88 & 1.97\\
  2M17292082-2126433 & \phantom{-} 3983.4 & \phantom{-} 0.78 & \phantom{-} -1.27 & 2.59\\
  2M17293482-2741164 & \phantom{-} 4143.5 & \phantom{-} 1.03 & \phantom{-} -1.25 & 1.85\\
  2M17323787-2023013 & \phantom{-} 3865.7 & \phantom{-} 1.03 & \phantom{-} -0.85 & 1.94\\
  2M17330730-2407378 & \phantom{-} 4042.5 & \phantom{-} 0.25 & \phantom{-} -1.87 & 1.88\\
  2M17341796-3905103 & \phantom{-} 4163.5 & \phantom{-} 1.42 & \phantom{-} -0.89 & 1.84\\
  2M17342067-3902066 & \phantom{-} 4380.4 & \phantom{-} 1.40 & \phantom{-} -0.90 & 1.99\\
  2M17503065-2313234 & \phantom{-} 3819.4 & \phantom{-} 0.98 & \phantom{-} -0.88 & 2.1\\
  2M18023156-2834451 & \phantom{-} 3617.4 & \phantom{-} 0.42 & \phantom{-} -1.19 & 3.02\\
  2M18143710-2650147 & \phantom{-} 4240.5 & \phantom{-} 1.30 & \phantom{-} -0.91 & 1.97\\
  2M18150516-2708486 & \phantom{-} 3833.4 & \phantom{-} 1.0  & \phantom{-} -0.82 & 2.14\\
  2M18344461-2415140 & \phantom{-} 4294.5 & \phantom{-} 1.09 & \phantom{-} -1.41 & 1.83\\

\hline
\noalign{\smallskip}
\hline 
\end{tabular}
\end{table}

\section{Analysis}

We have initially adopted the calibrated stellar parameters effective temperature
T$_{\rm eff}$, gravity log~g, metallicity [Fe/H] and microturbulence velocity v$_{t}$ from APOGEE DR16 - we point out that the calibrated parameters give very different element
abundances, and should not be used for such aims.
In fact the results from the  APOGEE Stellar Parameters and Chemical
Abundances Pipeline (ASPCAP) (Garc\'{\i}a-P\'erez et al. 2016) are obtained for the reported non-calibrated spectroscopic
stellar parameters.
We then adopted these non-calibrated stellar parameterers
from DR17, since we became aware that these are
obtained from a spectroscopic solution that minimizes the errors
in 7 dimensions (T$_{\rm eff}$,log~g, [Fe/H], v$_{t}$, [$\alpha$/Fe],
[C/Fe], [N/Fe]).

For this reason we proceeded with all the rederivation of abundances with
the DR17 non-calibrated parameters.
These stellar parameters are reported in Table \ref{param58_new},
and they are the final parameters adopted.

The abundances were determined by comparing the observed spectra with the synthetic ones. 
The synthetic spectra calculations are carried out with the code PFANT\footnote{The code is available at http://trevisanj.github.io/PFANT.}, as described in Barbuy et al. (2018b). This code is an update of the original FANTOM or ABON2 Meudon code by M. Spite. Each model atmosphere was interpolated in the MARCS grids (Gustafsson et al. 2008).
 
 The atomic line list employed is that from the APOGEE collaboration (Smith et al. 2021).
Molecular electronic transition lines of CN A$^{2}\Pi$-X$^{2}\Sigma$, vibration-rotation
CO X$^{1}\Sigma^{+}$, OH X$^{2}\Pi$ and TiO $\phi$-system b$^{1}$$\Pi$-d$^{1}$$\Sigma$ 
lines were included.  The line lists for CN were made available by
S. P. Davis, the CO line lists were adopted from
Goorvitch (1994), and the OH are from Goldman et al. (1998). For TiO
the line list is from Jorgensen (1994).
More details on CN, CO, OH and TiO molecular lines are given in
Mel\'endez \& Barbuy (1999), Mel\'endez et al. (2001, 2002, 2003), 
Schiavon \& Barbuy (1999) and Barbuy et al. (2018b).

 The atomic lines analysed initially were selected from 
 Smith et al. (2021), Shetrone et al. (2015),  Ce II lines identified by
 Cunha et al. (2017), and lines of S I  identified by  Fanelli et al. (2021).  
 Lines of Nd II (Hasselquist et al. 2016)  and Yb II (Smith et al. 2021) were not studied.
In Table \ref{linelist} are reported the lines that we verified in the
spectra of the 58 sample stars. 

For the moderately metal-poor sample stars, some of the lines indicated
in the articles above are not suitable, and
in a few cases we have added other lines that we identified as suitable
for the stellar parameters of the sample stars. The lines are discussed in detail below.
In the present work we adopt the ASPCAP abundances of Mg, Si, Ca and revise the C, N, O, and Ce abundances;
we also verified Ti lines and some comments are given, but the abundances are not used, given
conflicting results from different lines.
Other elements such as Na, Al and iron-peak elements will be analysed elsewhere.

We identified and fitted the studied lines in the reference stars Arcturus and
$\mu$ Leo, in order to check if the lines are well reproduced in these stars, and therefore
reliable for deriving abundances in the sample stars.
For the reference star Arcturus, we used the Hinkle et al. (1995) 
atlas, 
and for the metal-rich reference giant star $\mu$ Leo
a spectrum from APOGEE was used.
 The adopted stellar parameters for Arcturus and $\mu$ Leo
 are from Mel\'endez et al. (2003) and Zoccali et al. (2006) plus Lecureur et al. (2007), respectively.

Table \ref{sol} reports abundances in the Sun,
Arcturus and $\mu$ Leo. For the Sun they are from
a) Grevesse et al. 1996, 1998, adopted, 
b) Grevesse et al. (2015), Scott et al. (2015a,b), c) Lodders et al. (2009). 
For Arcturus, the abundances are from 
Mel\'endez et al. (2003), Lecureur et al. (2007), Ram\'{\i}rez \& Allende Prieto (2011), Barbuy et al. (2014), and Smith et al. (2013).
For $\mu$ Leo, the abundances are from , 
Gratton \& Sneden (1990), Smith \& Ruck (2000), Lecureur et al. (2007),
 Barbuy et al. (2015), Smith et al. (2013) or present fits,
using the observed spectrum by Lecureur et al. (2007) in the optical.

According to Ashok et al. (2021), and Nidever et al. (2015)
the average resolution of the APOGEE observations  
is R $\approx$ 22,500 based on a direct-measured FWHM of
$\sim$0.7Å, with 10-20\% variations seen across the wavelength range.
We have employed a typical FWHM=0.70 {\rm \AA}, but to fit better different lines we varied the FWHM values from 0.60 to 0.75, from the lowest to the highest wavelengths. Note that the FWHM varies from fibers to fibers and with a fiber with wavelength.


\begin{table*}
\small
\caption[4]{\label{linelist}
Line list. log~gf from VALD3 linelist (Piskunov et al. 1995, Ryabchikova et al. 2015), 
Kurucz (1993) and
NIST (Martin et al. 2002). The log~gf values for CeII lines are from Cunha et al. (2017).}
\begin{tabular}{lcccccccccccccc}
\hline\hline
\noalign{\smallskip}
\hbox{Species} & \hbox{$\lambda$} & \hbox{$\chi_{ex}$}  &\hbox{log~gf} &\hbox{log~gf} &\hbox{log~gf} & Notes & \\
& \hbox{(\AA)} &\hbox{(eV)} &\hbox{(VALD3)} & \hbox{(Kurucz)} & \hbox{(NIST)} & &   \\ 

\noalign{\smallskip}
\hline
\noalign{\smallskip}
\hbox{SiI} & 15361.161 & 5.954 & -1.925 & -1.990 & -1.710 & & \\
& 15376.831 & 6.222 & -0.649 & -0.290 & --- &  & \\
& 15833.602 & 6.222 & -0.168 & -0.660 & -0.078 & Apogee gap & \\
& 15960.063 & 5.984 & 0.107 & 0.130 & 0.197 &  & \\
& 16060.009 & 5.954 & -0.566 & -0.440 & -0.429 &  & \\
& 16094.787 & 5.964 & -0.168 & -0.110 & -0.078 &  & \\
& 16215.670 & 5.964 & -0.665 & -0.990 & -0.575 &  & \\
& 16680.770 & 5.984 & -0.140 & -0.500 & -0.090 &  & \\
& 16828.159 & 5.984 & -1.102 & -1.390 & -1.012 &  & \\
\hbox{CaI} & 16197.075 & 4.535 & 0.089  & 0.638 & --- & & \\
           & 16204.087 & 4.535 & -0.627  & -0.111 & --- &  & \\
\hbox{TiI} & 15543.756 & 1.879 & -1.120 & -1.273 & -1.080 &  & \\ 
& 15602.842 & 2.267 & -1.643 & -1.544 & --- &  & \\ 
& 15698.979 & 1.887 & -2.060 & -2.218 & -2.020 &  & \\ 
& 15715.573 & 1.873 & -1.250 & -1.359 & -1.200 & & \\
& 16635.161 & 2.345 & -1.807 & -2.178 & --- &  & \\
\hbox{CeII} & 15277.610 & 0.609  & -1.94 & --- & --- & too faint & \\
& 15784.750 & 0.318 & -1.54 & --- & --- &  & \\
& 15829.830 & 0.320  & -1.80 & --- & --- & Apogee gap & \\
& 15958.400 & 0.470 & -1.71 & --- & --- &  & \\
& 15977.120 & 0.232 & -2.10 & --- & --- & weak line strongly blended & \\
& 16327.320 & 0.561 & -2.40 & --- & --- &  & \\
& 16376.480 & 0.122 & -1.79 & --- & --- &  & \\
& 16595.180 & 0.122 & -2.19 & --- & --- &  & \\
& 16722.510 & 0.470 & -1.65 & --- & --- & & \\
\noalign{\hrule\vskip 0.1cm} 
\hline                  
\end{tabular}
\end{table*}   

\begin{table*}
\begin{flushleft}
\small
\caption{\label{sol}
Solar abundances from (1) Grevesse et al. (1996, 1998) (adopted);
(2) Steffen et al. (2015); (3) Scott et al. (2015a,b); (4) Grevesse et al. (2015); 
 (5) Lodders et al. (2009);
 Arcturus abundances from: 
(6) Ram\'{\i}rez \& Allende Prieto (2011),
(7) McWilliam et al. (2013),  (8) Lecureur et al. (2007),
 (9) Barbuy et al. (2014), (10) Smith et al. (2013); (11) Cunha et al. (2017)
$\mu$ Leo abundances from: (10) Smith et al. (2013); (12) Gratton \& Sneden (1990),
 13: Barbuy et al. (2015), (14) Van der Swaelmen et al. (2016);
(15) fits to the optical spectrum of $\mu$ Leo.} 
\begin{tabular}{lcccccccccccccc}
\hline\hline
\noalign{\smallskip}

\hbox{El.} & \hbox{Z} & \multispan3  \hbox{log $\epsilon(X)_{\odot}$}  &&[X/Fe] &\hbox{log $\epsilon(X)$} & &
[X/Fe] & \multispan2 \hbox{log $\epsilon(X)$} & \\

&  & \multispan3  Sun && \multispan2   \hbox{\rm Arcturus}  && \multispan3  \hbox{\rm $\mu$ Leo} &  \\
 && \multispan3       && \multispan2   \hbox{adopted} && \multispan3   \hbox{adopted} & \\
\noalign{\smallskip}
\hline
\noalign{\smallskip}
\noalign{\hrule\vskip 0.1cm}
\hbox{Fe} & 26  & 7.50 &  ~7.50 & ~7.50  && -0.54  &6.96  && +0.30  & 7.80 & 7.76 &\\
\hbox{C} & 6       &8.55[1]  &  ---     & ~8.39[5] &&-0.22[8]    & ~7.79     &&-0.3[10]   & 8.55 & 8.52 & \\
\hbox{N} & 7       &7.97[1]  &  ---     & ~7.86[5] &&+0.22[8]    & ~7.65     &&+0.45[10]  & 8.72 & 8.71 & \\
\hbox{O} & 8       &8.76[2]  &  ---     & ~8.73[5] && 0.39[9]     & ~8.62   && +0.0[10]  & 9.06 & 9.05 &\\
\hbox{Na} & 11     &6.33[1]  &  ~6.21[3]  & ~6.30[5] && 0.11[6]   & ~5.90    &&+0.50[8]   & 7.13 &  --- &\\
\hbox{Mg} & 12     &7.58[1]  &  ~7.59[3]  & ~7.54[5] && 0.37[6]   & ~7.41   &&-0.03[10]  & 7.85 & 7.85 & \\
\hbox{Al} & 13     &6.47[1]  &  ~6.43[3]  & ~6.47[5] && 0.37[7]   & ~6.30   &&+0.13[10]  & 6.90 & 6.90 & \\
\hbox{Si} & 14     &7.55[1]  &  ~7.51[3]  & ~7.52[5] && 0.33[6]   & ~7.34   &&-0.10[10]  & 7.75 & 7.76 &\\
\hbox{Ca} & 20     &6.36[1]  &  ~6.32[3]  & ~6.33[5] && 0.11[6]   & ~5.93   &&-0.04[10]  & 6.62 & 6.62 &\\
\hbox{Sc} & 21     &3.17[1]  &  ~3.16[3]  & ~3.10[5] && 0.23[6]   & ~2.86    &&+0.10[11]  & 3.57 & ---  &\\
\hbox{Ti} & 22     &5.02[1]  &  ~4.93[3]  & ~4.90[5] && 0.26[7]   & ~4.74   &&+0.10[10]  & 5.42 & 5.40 &\\
\hbox{V} & 23      &4.00[1]  &  ~3.89[3]  & ~4.00[5] && 0.12[7]   & ~3.58   &&+0.03[12]  & 4.33 & 4.18 &\\
\hbox{Cr} & 24     &5.67[1]  &  ~5.62[3]  & ~5.64[5] && -0.05[6 ] & ~5.08   &&-0.01[12]  & 5.96 & 6.14 &\\
\hbox{Mn} & 25     &5.39[1]  &  ~5.42[3]  & ~5.37[5] && -0.14[7]  & ~4.71   &&+0.00[12]  & 5.69 & 5.79 &\\
\hbox{Co} & 27     &4.92[1]  &  ~4.93[3]  & ~4.92[5] && +0.09[7]  & ~4.49   &&+0.00[12]  & 5.22 & 5.23 &\\
\hbox{Ni} & 28     &6.25[1]  &  ~6.20[3]  & ~6.23[5] && 0.06[6]   & ~5.77   &&+0.05[10]  & 6.60 & 6.60 &\\
\hbox{Cu} & 29     &4.21[1]  &  ~4.19[4]  & ~4.21[5] && -0.26[10] & ~3.55   &&-0.10[10]  & 4.41 & 4.41 &\\
\hbox{Zn} & 30     &4.60[1]  &  ~4.56[4]  & ~4.62[5] && +0.18[6]  & ~4.26   &&-0.10[13]  & 4.80 & --- &\\
\hbox{Y} & 39      &2.24[1]  &  ~2.21[4]  & ~2.21[5] && -0.30[9]  & ~1.40    &&+0.04[14]  & 2.58 & ---  &\\
\hbox{Zr} & 40     &2.60[1]  &  ~2.59[4]  & ~2.58[5] && -0.28[7]  & ~1.78   &&+0.10[12]  & 3.00 & ---  &\\
\hbox{Ba} & 56     &2.13[1]  &  ~2.25[4]  & ~2.17[5] && -0.30[9]  & ~1.29   &&+0.10[14]  & 2.53 & ---  &\\
\hbox{La} & 57     &1.22[1]  &  ~1.11[4]  & ~1.14[5] && -0.30[9]  & ~0.38    &&-0.37[14]  & 1.15 & ---  &\\
\hbox{Ce} & 58     &1.55[1]  &  ~1.58[4]  & ~1.61[5] && -0.45[11]  & ~0.99    &&-0.37[14]  & 1.15 & ---  &\\
\hbox{Eu} & 63     &0.51[1]  &  0.52[4]   & ~0.52[5] && 0.23[7]   & ~0.20    &&-0.14[14]  & 0.67 & --- &\\
\noalign{\hrule\vskip 0.1cm} 
\hline                  
\end{tabular}
\end{flushleft}
\end{table*}

\subsection{C, N, O abundances}

The abundances of C, N and O are derived from CN, OH and CO molecular lines. 
They are interdependent due to the molecular dissociative equilibrium. Since the molecular lines are spread all over the
spectra, these abundances are derived first, and they are reported in
Table \ref{cnodr17noncal}.

Computing synthetic spectra employing the PFANT code described in Barbuy et al. (2018b), we have derived C, N, O abundances in two ways:\newline
\indent
{\it method a):} in the region 15144-16896 {\rm \AA}, first we derive the O abundances
by analysing the molecular lines of OH. Some of the most prominent OH
lines in this region are at: 15264.60, 15266.160, 15278.516, 15281.045,
15719.687, 15893.524, 16074.151, 16662.187, 16872.265, 16895.164 {\rm \AA}.
These lines are the most sensitive to oxygen variation in the APOGEE sample.
We derive C abundances by analysing the CO molecular lines, but there are not
many strong CO lines in the range of 15100-17000 {\rm \AA}. In our sample,
the strongest lines of CO, used to measure C abundances, are at 15983.214,
15985.598, 15990.420, 16016.081 {\rm \AA}. Next we see how the CN lines
change when we modify Nitrogen. The most sensitive lines of CN are at
15162.648, 15222.382 {\rm \AA}. There are many CN lines in the region
we are working with (especially in the range 15522-15600 {\rm \AA}),
but most of them are too shallow to give reliable abundance measurements.

\indent
{\it method b):} a derivation of CNO abundances using  the region
15525-15595 {\rm \AA}, where there are clear lines of  OH, and a clear
bandhead of CO, as well as lines of CN, although less conspicuous,
as done for example in Barbuy et al. (2021a) for Phoenix
spectra  that were observed in this region only.
For these calculations a FWHM=0.60 was adopted which is suitable
for the wavelength region in question.

\begin{figure*}
	\includegraphics[angle=0,width=14cm]{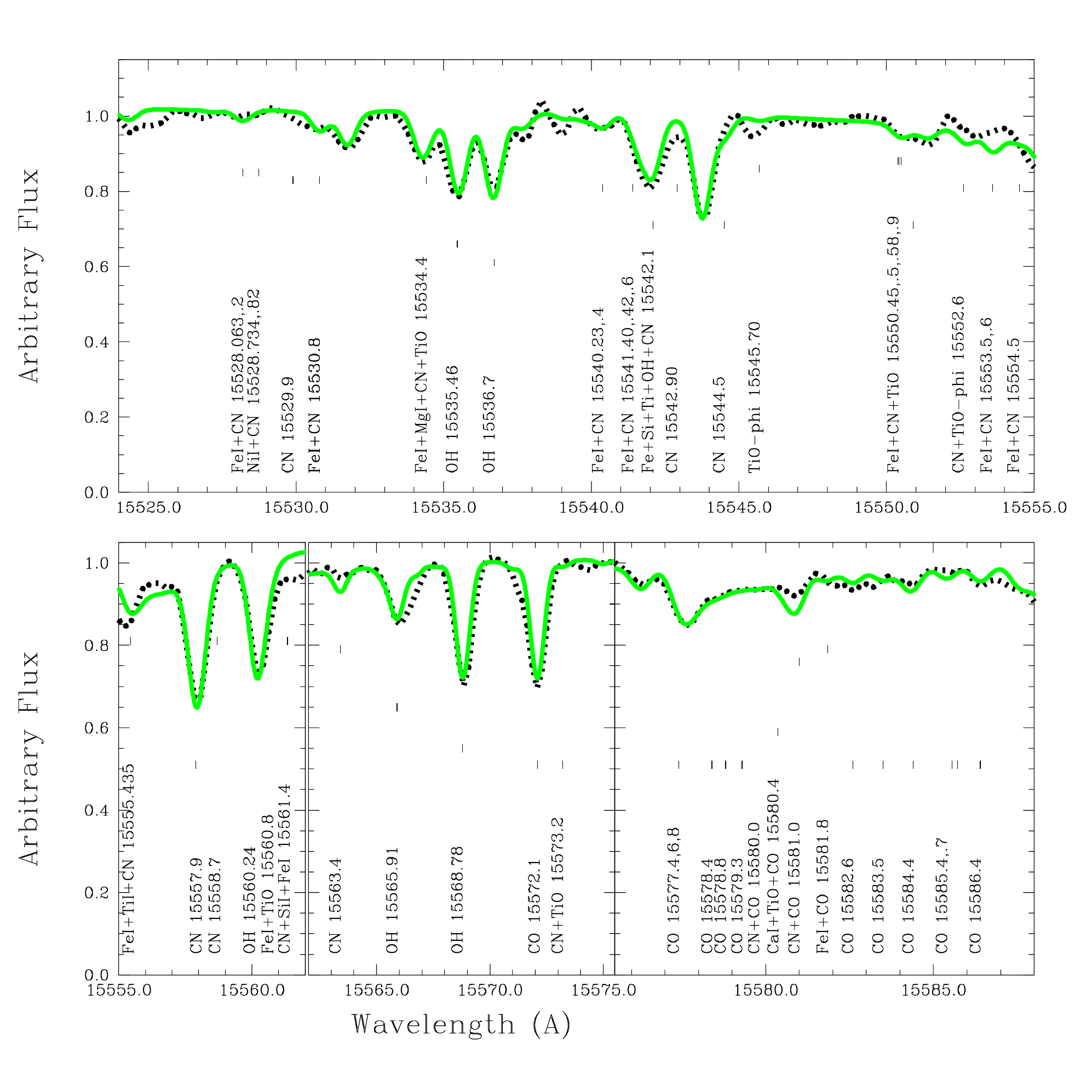}
        \caption{Star 2M17382504-2424163: Observed spectrum (black dotted) and
          synthetic spectrum computed with
        [C,N,O/Fe] = -0.20, 0.30, 0.40 (green).}
    \label{c17}
\end{figure*}

This is illustrated in Figure \ref{c17} 
   for star 2M17382504-2424163. 
   Note the clear OH lines at 15535.46, 15565.91, 15566.78,  and
clear CO bandhead at 15577.4 {\rm \AA}.

We concluded that both methods a) and b) give very similar results
within $\pm$0.1dex.

A verification of these CNO abundances was carried out by
fitting lines along all the spectra, in particular the lines
of CO 15600.74, 15612.5, 15667.55 {\rm \AA},
where only for four stars the C abundance was decreased by
-0.05 to -0.10 (stars 
2M17330695-2302130, 2M18050663-3005419, 2M18125718-2732215 and 
2M18344461-2415140), and for the others the fits were very
satisfactory.

We then proceeded with the verification of the OH lines:
OH 15130.921, 15266.168, 15281.052, 15409.172, 15568.78,
15651.896, 15719.696, 15755.522 {\rm \AA}, and CN 5181.277, 15298.487,
15308.893, 15318.74, 15337.959, 15341.508, 15432.811,
15447.095, 15466.235, 15481.868, 15530.776, 15684.088,
15737.445 {\rm \AA}. 
Only for star 
2M18023156-2834451 we increased the oxygen abundance by 0.05 dex,
noting that its spectra shows larger lines than the others, needing
a higher spectral convolution to be fitted.

\begin{figure*}
	\includegraphics[angle=0,width=14cm]{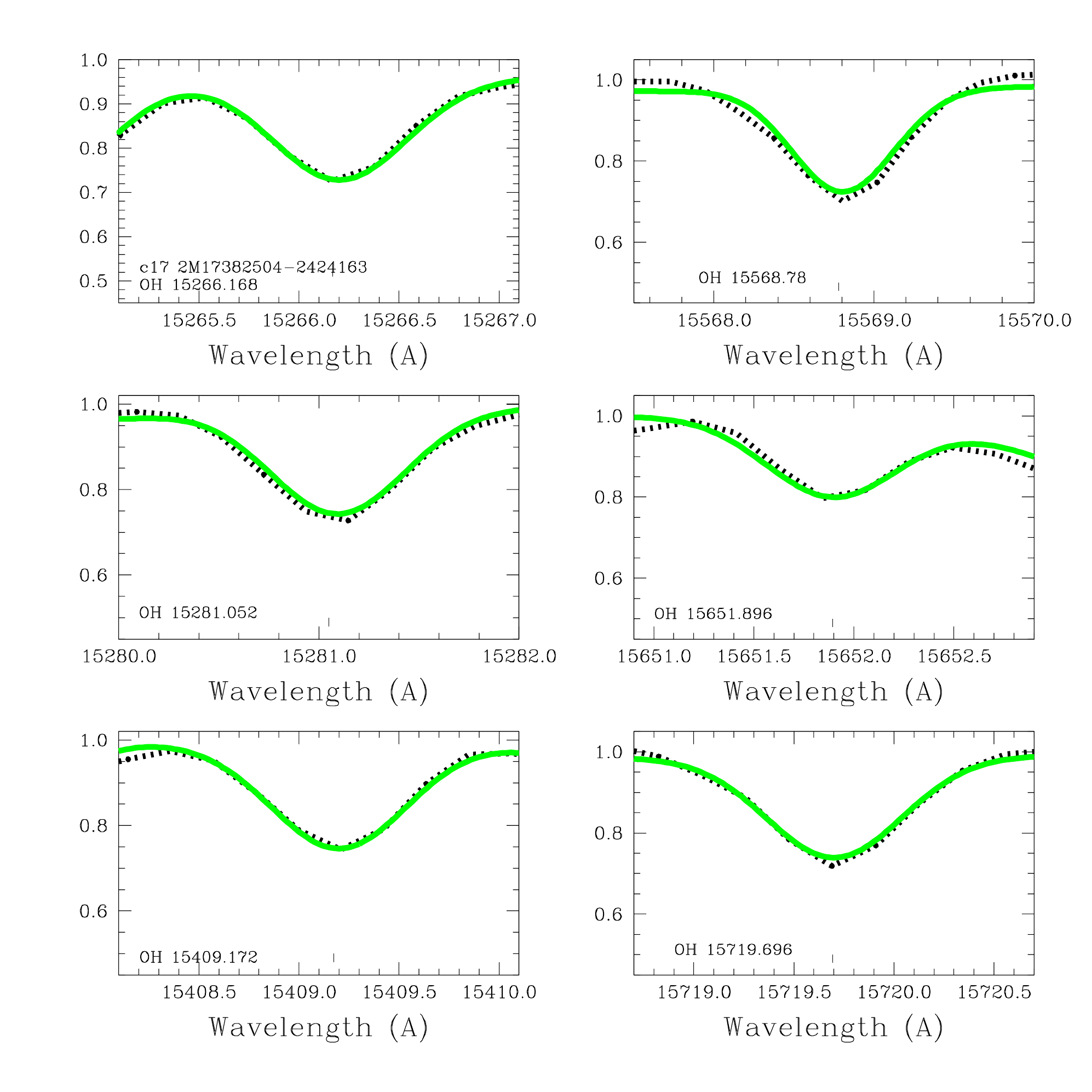}
        \caption{Star 2M17382504-2424163: Selected OH lines. Observed spectrum (black dotted) and synthetic spectrum computed with [O/Fe] = 0.40 (green).}
    \label{c17oh}
\end{figure*}

\begin{figure*}
	\includegraphics[angle=0,width=14cm]{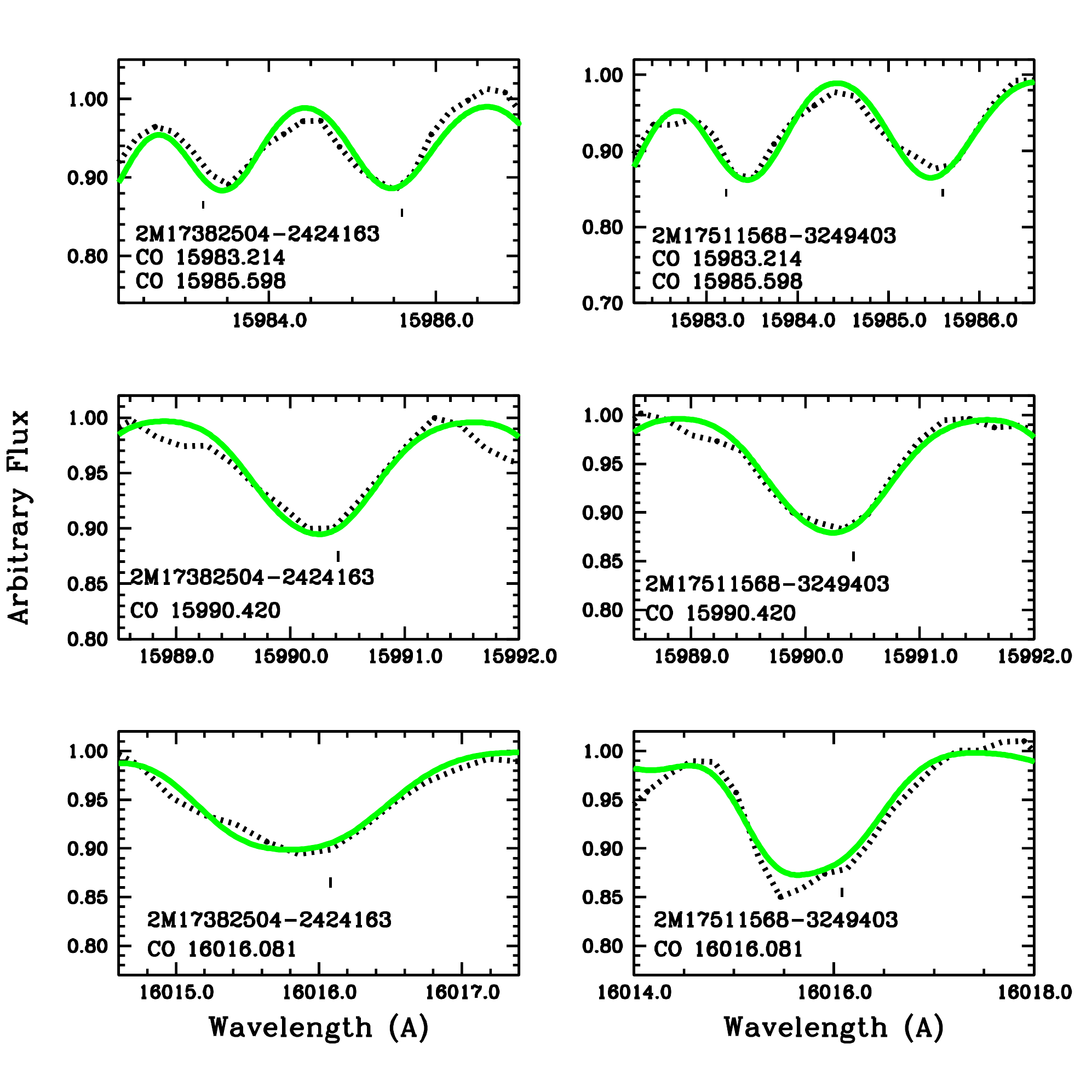}
            \caption{Stars 2M17382504-2424163 and 2M17511568-3249403: Selected CO lines. Observed spectra (black dotted) and
          synthetic spectra computed with
      [C/Fe] = -0.2 for both stars, and [O/Fe] = 0.40 and 0.38 respectively (green).}
    \label{c1718co}
\end{figure*}

Fits are shown for selected OH lines for star 
2M17382504-2424163 
in Figure \ref{c17oh}, 
and CO lines in Figure 
 \ref{c1718co} for stars 2M17382504-2424163 and 2M17511568-3249403.

\begin{figure*}
	\includegraphics[angle=0,width=14cm]{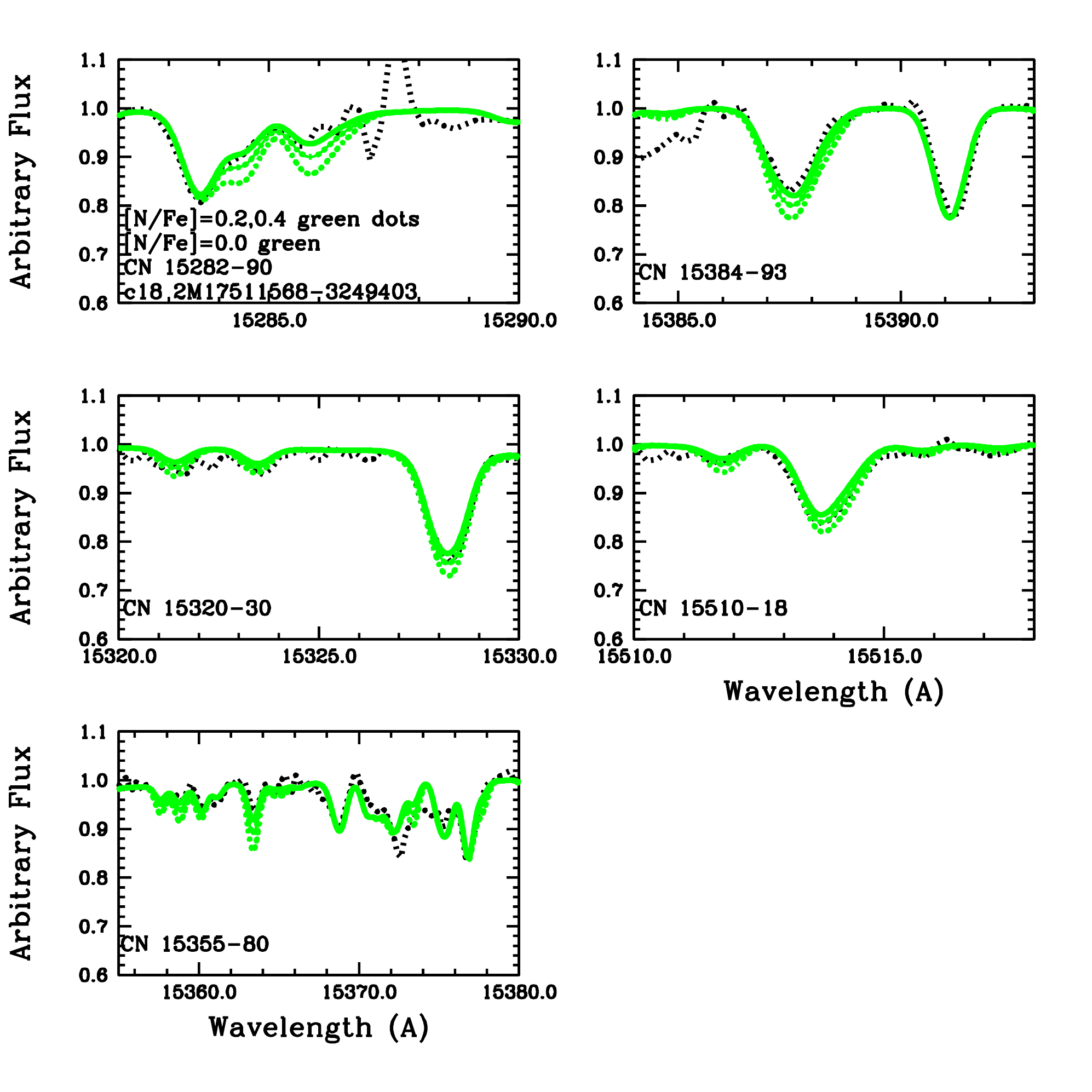}
        \caption{Star 2M17511568-3249403: regions containing CN lines. Observed spectra (black dotted) and
          synthetic spectra (green) computed with
      [C/Fe] = -0.2, [N/Fe] = 0.0, [O/Fe] = 0.38; green dotted lines correspond
      to calculations with [N/Fe]=0.2, 0.4.}
    \label{c18cn}
\end{figure*}

Regions of CN lines are verified, using wavelength
regions indicated by Fern\'andez-Trincado et al. (2020a,b)
for example. In Figure \ref{c18cn} are shown the fits to good CN lines.
Among these, the clearest CN feature is at 15387.6 {\rm \AA}, and its fits are compatible with the
C,N abundances from the 15283-15287, 15320-15330 and 15355-15380 {\rm \AA} regions.
The feature at 15514 {\rm \AA} is blended with  a CoI line and is less reliable.
The N abundances derived 
are confirmed for about half the stars, whereas for the other half
the N abundance was decreased by a mean of 0.2dex: this is not surprising
because the CN lines in the 15555$\pm$50 {\rm \AA} from method b) are all faint and/or
blended.
 Results from method b) above, together with these corrections, are
adopted for C, N, O abundances.

The results of our manual analysis differ  from the outputs of the ASPCAP pipeline for oxygen and, to a lesser degree, nitrogen. 
Our methods a) and b) give rather similar results to each other within
$\pm$0.05dex, and with oxygen abundances somewhat higher than those derived with ASPCAP,
that appear to be too low for bulge stars.
The uncertainties on the oxygen abundances were already discussed by J\"onsson et al. (2018) and
Zasowski et al. (2019). Our oxygen abundances as compared with the DR17 ones are compatible
within uncertainties, but with a trend to be higher.

In order to verify the reason for these differences, we carried
out the fit to the N-rich star 2M17480576-2445000 analysed by
Schiavon et al. (2017). With our method b) we have found that
[O/Fe]=0.4 instead of [O/Fe]=0.3, and [C/Fe]=-0.2, instead of
[C/Fe]=0.0, and on the other hand the N enhancement of
[N/Fe]=0.8 is confirmed. Given the interplay between the CNO trio
elements, it appears that the trend is to have somewhat lower C
and higher O, and not much of a change in N abundances, in comparing
our abundances with those from ASPCAP.

Note that none of the stars in our sample is N-enhanced, therefore they are good 
candidates to being similar to the first generation stars found in globular clustes.

\subsection{alpha-elements Mg, Si, Ca and Ti}


We analyse the abundances of the $\alpha$-elements Mg, Si, and Ca,
and the iron-peak element Ti.

{\it Magnesium, Silicon, Calcium and Titanium} 

Our fits with the original DR17 ASPCAP Mg abundances are in agreement with their results for Mg, Si and Ca. Our calculations are in LTE with
plane parallel models, as is adopted by the original ASPCAP method. 
The DR17 results for Mg and Ca correspond to calculations in non-LTE
(Osorio et al. 2020), and even so the compatibility is good for
these elements.


The SiI lines reported in Table \ref{linelist}
are all suitably reproduced with the ASPCAP Si
abundance, with the exception of line SiI 15261.161 that is too
shallow in the sample stars. Si abundance appears to be among the
best determined ones by ASPCAP, together with Mg.

The four CaI lines listed in Smith et al. (2013) 
and that are use in ASPCAP, namely
16136.8, 16150.8, 16155.2, 16157.4 {\rm \AA} (see also J\"onsson et al. 2018) are faint in the
sample stars, and they are not fitted with the ASPCAP Ca abundance; instead they would need an extra 0.2dex in Ca abundance to be fitted.
In Table \ref{linelist} we include another two lines of CaI
that we were able to identify as suitable for the metallicity of our stars: CaI 16197.075, 16204.087 {\rm \AA}. 
The CaI 16197.075 is well fitted in about half the stars, whereas in others it show blends, and finally the CaI 16204.087 {\rm \AA}
 line is well fitted with the Ca abundance from ASPCAP.
A FWHM=0.65 fits better the lines.
In conclusion, we adopted the ASPCAP Ca abundances, relying on the results for the CaI 16204.087 {\rm \AA} line.

{\it Titanium:} 
Among the 5 lines studied, only TiI 15543.756 {\rm \AA} line is well fit  in essentially
all stars. TiI 15698.979 {\rm \AA} tends to give the same value, but it is located in a blend with several other lines, with a difficult continuum placement.
TiI 15715.753 {\rm \AA} tends to give either
the value from ASPCAP or requires a lower Ti abundance, whereas TiI 15602.842, 
and 16635.161 {\rm \AA} require higher values by about 0.3$\pm$0.2 dex to be fitted.
Because of the conflicting results from these different lines, and the fact that ASPCAP gives
[Ti/Fe]=0.0 for most stars, which is not compatible with the Si and Ca enhancements, we
preferred not to analyse the Ti abundances in the sample stars. 

Note that the lines TiI 15602.842,  and 16635.161
{\rm \AA}, that are only fitted with higher Ti abundances, have 
somewhat higher excitation potential than the other
3  inspected lines, and that means that there may be an effect of
effective temperature.

\subsection{s-process element Ce}

We used 6 CeII lines, among which
CeII 16722.510 {\rm \AA} line is  well fit to almost all stars, 
except for a few for which most
of the other lines are fit with a lower value than with the best line (case of 2M17173693-2806495), 
followed by CeII 15958.400 and 16595.180 {\rm \AA} lines, 
that are fit with the adopted
value for almost all stars.

CeII 15784.750 {\rm \AA} is fit for about half the stars, for a few would require
lower Ce abundances and about 1/3 of them would require higher Ce abundances;
 16327.320 {\rm \AA} is faint and is fit for about 1/3 of stars and 2/3 would require
 higher Ce abundances;
CeII 16376.480 {\rm \AA} would require higher values for about half the stars.

\begin{figure*}
	\includegraphics[width=17cm]{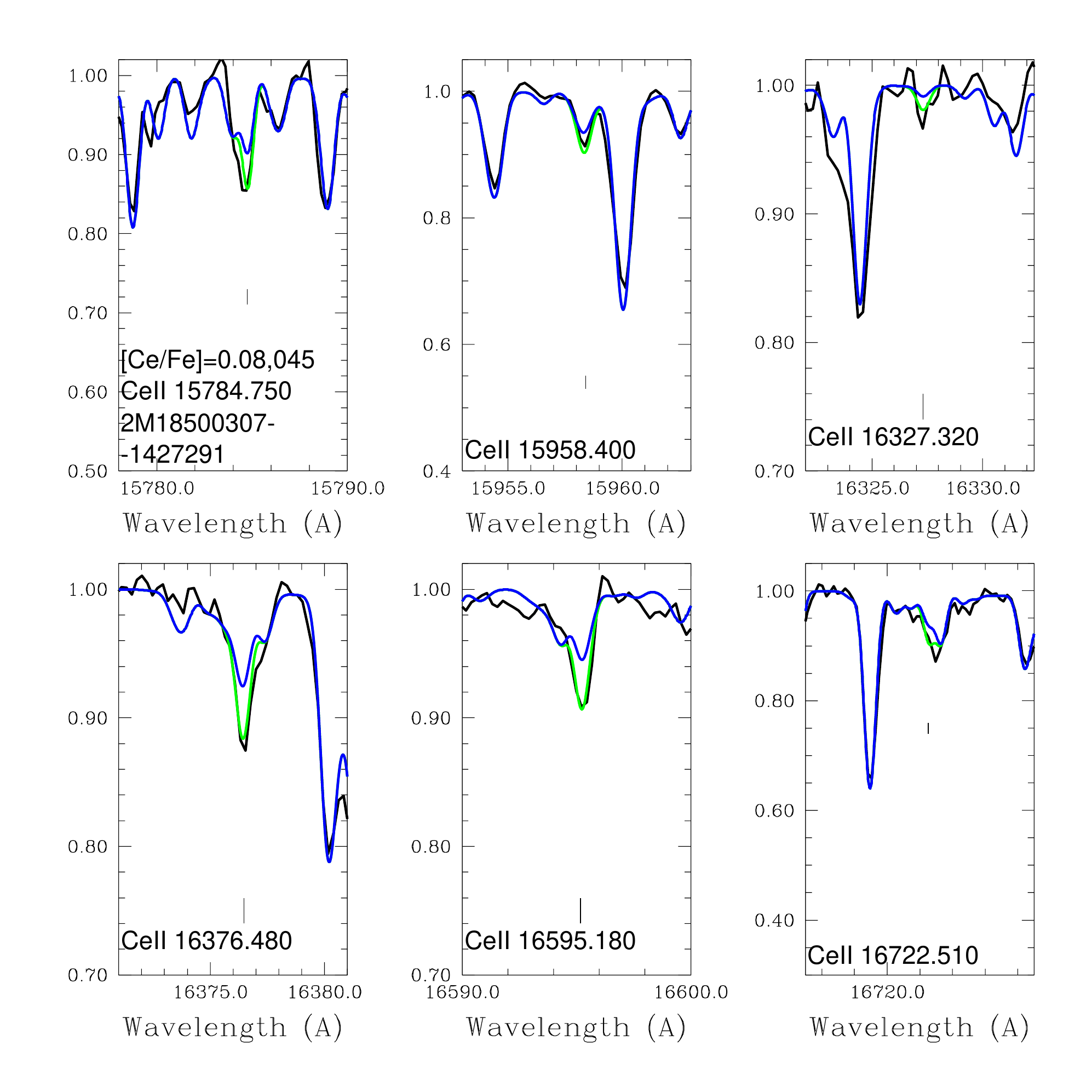}
    \caption{Star 2M18500307-1427291: fit to the 6 cerium lines.
    Observed spectrum: black; śynthetic spectra are: blue with original
    ASPCAP [Ce/Fe]=0.08 Ce abundance, green with final Ce abundance.}
    \label{b27cex}
\end{figure*}

In 8 cases all six lines can be considered well-fitted, as is the case of
star 2M18500307-1427291, shown in Figure \ref{b27cex}.

For the fit of the Ce lines, we adopted FWHM=0.75, which is suitable for the wavelength
of the lines. The revised values are systematically higher than those
resulting from ASPCAP (the fits to all stars are available under request).

DR17 used three Ce II windows covering the lines 15784, 16376, and 16595
{\rm \AA}.
These are the stronger lines among the six that we used, and for
all the sample stars it is clear that, from these 3 lines,
a higher Ce abundance is needed.

We note that due to uncertainties in the Ce abundances in DR17 the APOGEE
team has released internally to the collaboration
a value added catalogue with revised abundances (Hayes et al. in
preparation). This catalogue will be public to the community
in a few months.

As for Nd we found that the lines are not suitable for analysis, from fits to them in the reference stars Arcturus and $\mu$ Leo, therefore we disregarded this element in
the present analysis.

\section{Discussion}


The $\alpha$-element abundances in bulge stars provide us with a constraint on the formation history of its stellar populations: the formation timescale. In other words, a mean [$\alpha$/Fe]$\sim$0.5 in halo and bulge metal-poor stars of [Fe/H]$\simless-$1.0 indicates a fast chemical enrichment at early times, dominated by supernovae type II (SNII)
(e.g. Woosley \& Weaver 1995, hereafter WW95), whereas a lower [$\alpha$/Fe] implies a slower enrichment, allowing supernovae type Ia to contribute to the enrichment of iron.

Moreover, as recently shown by Miglio et al. (2021) for a sample of Kepler stars with APOGEE spectra,
stars with [$\alpha$/Fe]$>$0.2 are all very old. The same probably applies to the present sample.

\subsection{Oxygen  and magnesium}
\begin{figure*}
	\includegraphics[angle=0,width=17cm]{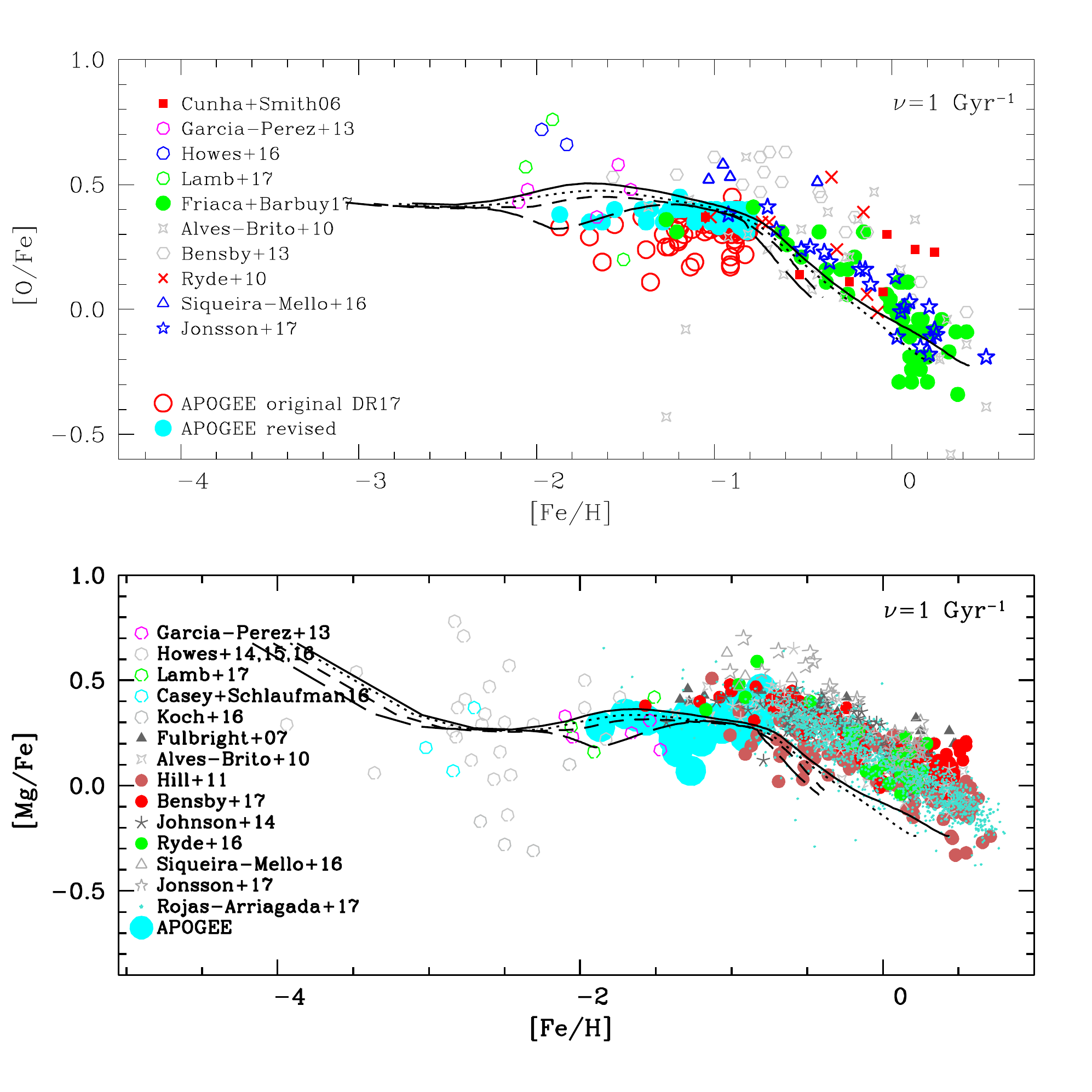}
        \caption{[O/Fe] vs. [Fe/H] (upper panel) and [Mg/Fe] vs. [Fe/H]
        (lower panel), for literature bulge field stars and the APOGEE
        abundances (original and revised in the case of oxygen) for the 58
        sample stars.
        Symbols: grey 4-pointed stars: Alves-Brito et al. (2010);
red filled circles: Bensby et al. (2013);
red filled circles: Bensby et al. (2017);
grey open pentagons: Casey \& Schlaufman (2015);
strong-grey filled triangles: Fulbright et al. (2007);
magenta open pentagons: Garc\'{\i}a-P\'erez et al. (2013);
red filled squares: Cunha \& Smith (2006);
indianred filled circles: Hill et al. (2011);
grey open pentagons: Howes et al. (2016);
grey stars: Johnson et al. (2014);
grey 5-pointed stars: J\"onsson et al. (2017);
grey open pentagons: Koch et al. (2016);
green open pentagons: Lamb et al. (2017);
red crosses: Ryde et al. (2010);
green filled circles: Ryde et al. (2016);
turquoise 5-pointed stars: Rojas- Arriagada et al. (2017);
grey open triangles: Siqueira-Mello et al. (2016);
blue open circles: APOGEE original;
cyan filled circles: final abundances for the 58 APOGEE sample stars.
The oxygen abundances are normalized in terms of adopted solar
abundances as explained in Fria\c ca \& Barbuy (2017).
Chemodynamical evolution models from Fria\c ca \& Barbuy (2017) with formation timescale of 1 Gyr,
for several radii, are overplotted:
$r<0.5$ kpc (solid lines), $0.5<1$ kpc (dotted lines),  $1<r<2$ kpc
(short-dashed lines), $2<r<3$ kpc (long-dashed lines).}
\label{oxmg}
\end{figure*}

Oxygen is produced by helium and neon burning in hydrostatic phases
of the evolution of massive stars.
Magnesium, together with Aluminum, are produced in hydrostatic carbon
and neon burning (WW95). O and Mg are therefore the bona-fide alpha-elements
produced by massive stars and ejected by 
supernova type II (SNII) event. They are enhanced
relative to iron in old stellar populations, such as in bulge
stars. In Figure \ref{oxmg} (upper panel) are plotted the oxygen abundances
reported by the original APOGEE DR17 release, and the revised values
obtained as explained in Section 3. This Figure is readapted from that
in Barbuy et al. (2018a), taking into account only the literature higher-resolution
data (with a few exceptions) and data showing little abundance spread.
The literature data taken into account are
from Fria\c ca \& Barbuy (2017), that contains a revision
of the abundances from Zoccali et al. (2006) and 
Lecureur et al. (2007), Cunha \& Smith (20060, Alves-Brito et al. (2010), Fulbright et
al. (2007), only stars older than 11
Gyr from Bensby et al. (2013), Ryde et al. (2010) including
 a few of the same stars from  Zoccali et al. (2006),
J\"onsson et al. (2017) including reanalysed
23 stars from Zoccali et al. (2006), Lecureur et al.
(2007) and Fria\c ca \& Barbuy (2017),
Siqueira-Mello et al. (2016), and metal-poor stars from
Garc\'{\i}a-P\'erez et al. (2013), Howes et al. (2016) and Lamb et al. (2017).



Figure \ref{oxmg} (lower panel) gives [Mg/Fe] vs. [Fe/H] for
metal-poor stars from Garc\'{\i}a-P\'erez et al. (2013), Howes et al. (2016), Lamb et al. (2017), Casey \& Schlaufman (2015),
Koch et al. (2016), Fulbright et al. (2007), as corrected by
McWilliam (2016), Alves-Brito et al. (2010), Hill et al. (2011),
Bensby et al. (2017) for stars older than 8 Gyr,
Johnson et al. (2014), Ryde et al. (2010, 2016), Siqueira-Mello
et al. (2016), J\"onsson et al. (2017) and Rojas-Arriagada et al.
(2017).

Our fits show agreemeent with the APOGEE ASPCAP Mg abundances, and they are compatible with the Mg abundances of other samples
of bulge stars.
The different model lines in Figure \ref{oxmg} correspond to different
radii from the Galactic center.

\subsection{Chemodynamical evolution model for O and Mg}

We have compared the abundances derived from observations  with the predictions of chemodynamical evolution models for the bulge (Friaça \& Barbuy 2017), described as a classical spheroid. It is assumed a baryonic mass of 2$\times$10$^9$ M$_{\odot}$, and a dark halo mass $M_H$= 1.3$\times$10$^{10}$ M$_{\odot}$. One central parameter of the model is the specific star formation rate $\nu_{SF}$ (i.e. the inverse of the star formation time scale).

In the nucleosynthesis prescriptions of our model, we adopt the metallicity dependent yields from core-collapse supernovae (SNe II) from WW95, with some modifications following suggestions of Timmes et al. (1995). For low metallicities (Z<0.01Z$_{\odot}$), we included the yields  from high explosion-energy hypernovae (HNe) (Nomoto et al. 2013, and references therein). The type Ia supernovae yields are from Iwamoto et al. (1999)  – their models W7 (progenitor star of initial metallicity Z=Z$_{\odot}$) and W70 (zero initial metallicity). The yields for intermediate mass stars (0.8-8 M$_{\odot}$)) with initial Z=0.001, 0.004, 0.008, 0.02, and 0.4 come from van den Hoek \& Groenewegen (1997) (variable $\eta_{AGB}$ case).

As we can see from Figure \ref{oxmg}, the abundances derived here both for the oxygen (upper panel) and for the magnesium (lower panel) are well reproduced by the chemodynamical model with $\nu_{SF} =$ 1 Gyr$^{-1}$  (star formation time scale of 1 Gyr). Once more this suggests these objects to be very old.

\subsection{Silicon and calcium}

Si and Ca are mainly produced during the
explosive nucleosynthesis of SNII events (WW95; McWilliam 2016), with smaller contributions from supernovae type Ia (SNIa).

\begin{figure*}
\includegraphics[angle=0,width=17cm]{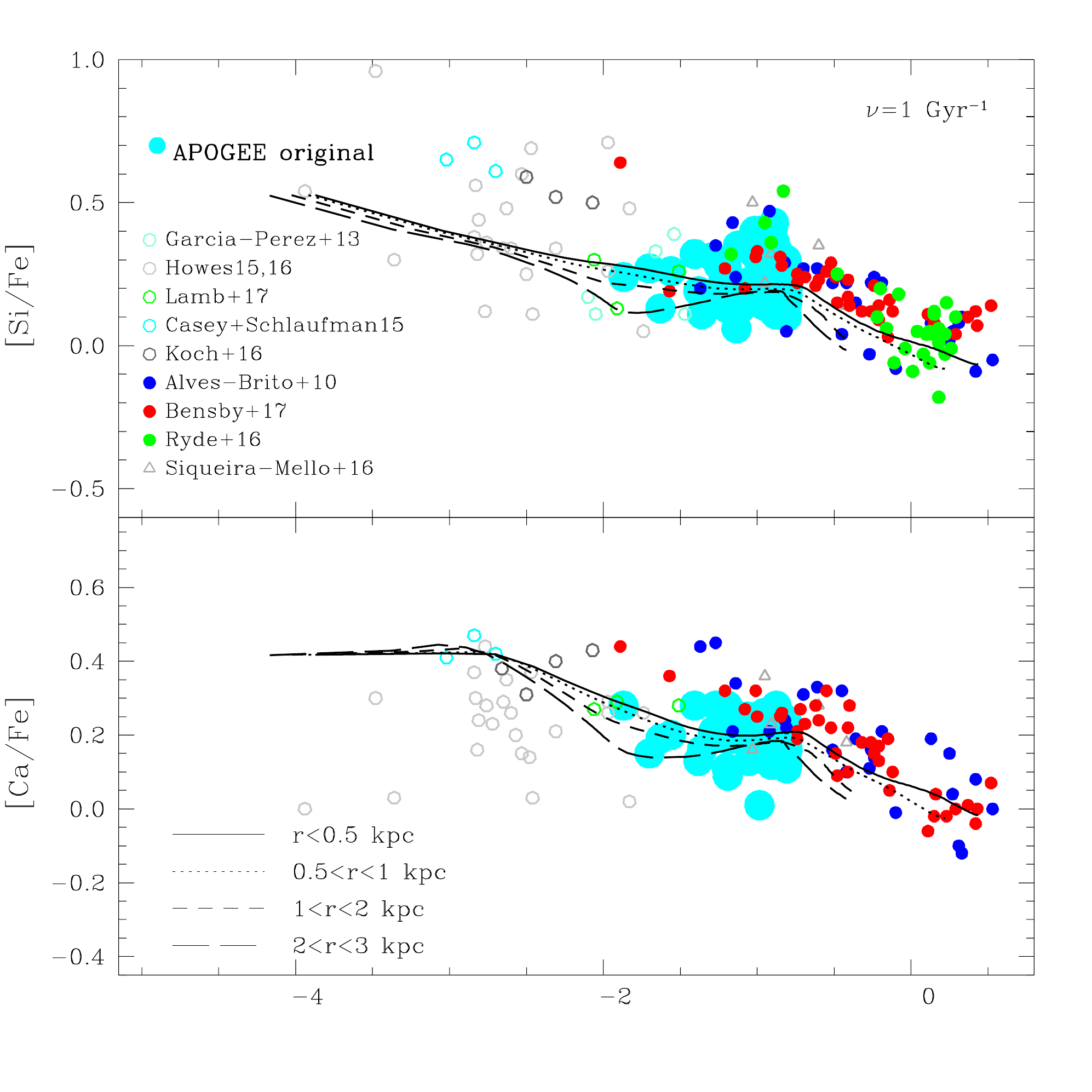}
\caption{[Si,Ca/Fe] vs. [Fe/H]  for literature bulge field stars and the APOGEE
abundances for the 58 sample stars.
Symbols: grey 4-pointed stars: Alves-Brito et al. (2010);
red filled circles: Bensby et al. (2017);
grey open pentagons: Casey \& Schlaufman (2015);
magenta open pentagons: Garc\'{\i}a-P\'erez et al. (2013);
grey open pentagons: Howes et al. (2016);
grey open pentagons: Koch et al. (2016);
green open pentagons: Lamb et al. (2017);
green filled circles: Ryde et al. (2016);
grey open triangles: Siqueira-Mello et al. (2016);
blue open circles: APOGEE original;
cyan filled circles: final abundances for the 58 APOGEE sample stars.
The lines are the predictions of the chemodynamical models of 
Fria\c ca \& Barbuy (2017) with a formation timescale of 1 Gyr for several radii.
}
\label{plotsica}
\end{figure*}

The $\alpha$-elements Si and Ca are plotted in Figure
\ref{plotsica} for the 58 sample stars
together with literature data from
Garc\'{\i}a-P\'erez et al. (2013), Howes et al. (2016), Lamb et al. (2017),
Casey \& Schlaufman (2015), Koch et al. (2016), Alves-Brito et al. (2010),
Bensby et al. (2017) for stars older than 8 Gyr,
Ryde et al. (2016) and Siqueira-Mello et al. (2016).

Figure \ref{plotsica} shows that a typical star formation time scale of 1 Gyr (the chemodynamical model with  $\nu_{SF} =$ 1 Gyr$^{-1}$) also explains the Si and Ca abundances found in the bulge.

From this figure we can conclude that there are no differences in the Si and Ca
abundances of the present confirmed in-situ samples of bulge stars, and previous samples in bulge regions,
for which no precise distances were available.

\subsection{Heavy element Ce}
\begin{figure}
\includegraphics[angle=0,width=9cm]{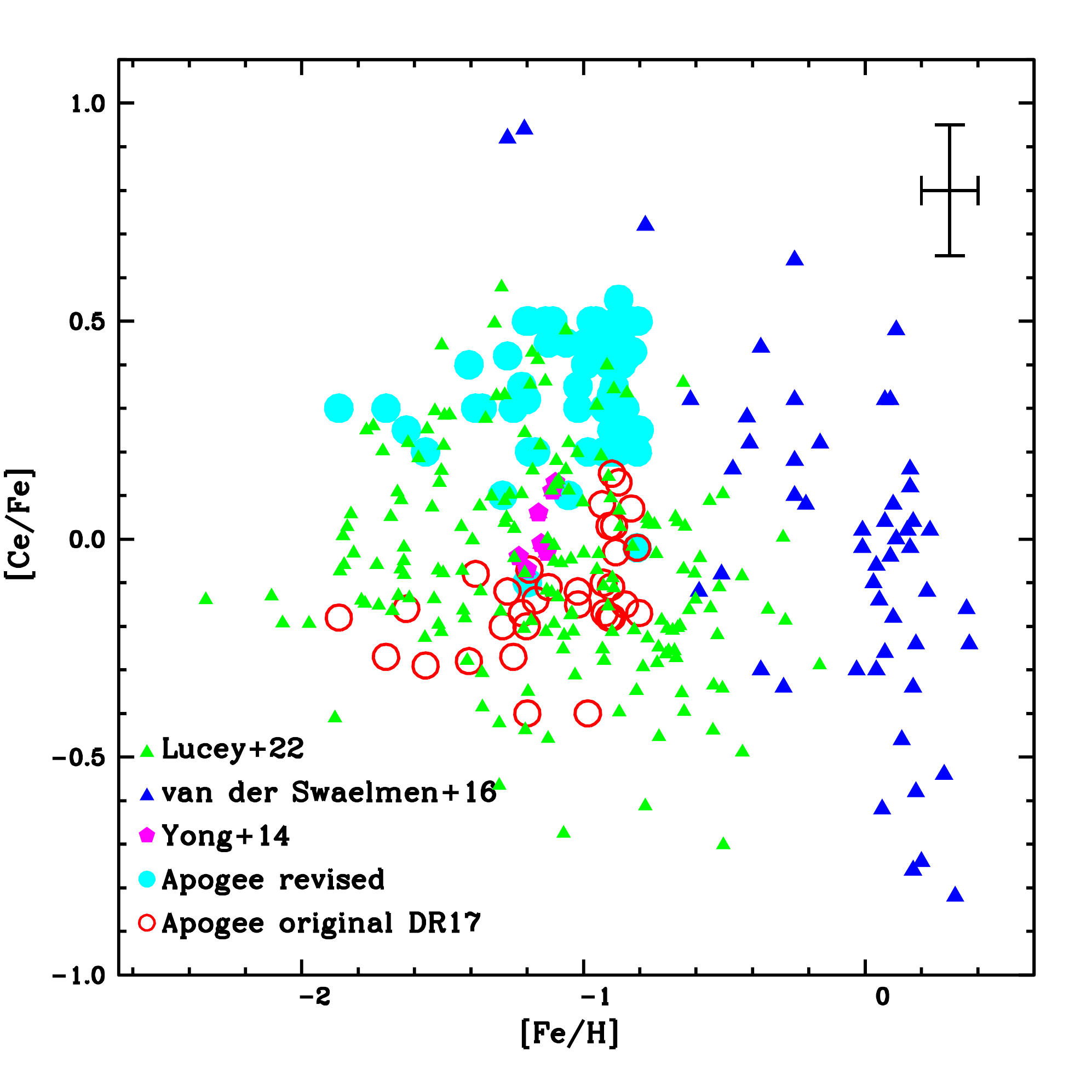}
\caption{[Ce/Fe] vs. [Fe/H]  for literature bulge field stars and the APOGEE
abundances for the 58 sample stars.
Symbols:
red filled triangles: van der Swaelmen et al. (2016);
green filled triangles: Lucey et al. (2022);
magenta pentagons: M62 from Yong et al. (2014);
open blue circles: APOGEE DR17 [Ce/Fe] values and
cyan filled circles: revised abundances for the 58 APOGEE sample stars.
Error bars are indicated in the right upper corner.
}
\label{plotcex}
\end{figure}

In Figure \ref{plotcex} are shown the revised Ce abundances
in contrast with the lower original DR17 APOGEE abundances, together
with results for M62 from Yong et al. (2014) and for field bulge
stars from van der Swaelmen et al. (2016) and Lucey et al. (2022).
As can be seen from the figure,
we have found that the sample stars are enhanced in Cerium, by about a mean value
of [Ce/Fe]$\sim$0.4. We find Ce enhancements relative to the DR17 and as well
to results from the COMBS survey by Lucey et al. (2022). Cleary further
investigation on Ce abundances in metal-poor bulge stars is needed.

Ce (Z=58, A=140) is essentially an element mostly formed by the
s-process, with a fraction of
0.186 as r-element and  0.814 as s-element
(Simmerer et al. 2004).
Ce appears to be overproduced in massive spinstars (Frischknecht et al. 2016). On the other hand,
since we do not have the ages of these stars, we cannot
exclude that the Ce enhancement could be due to a mass
transfer from a companion Asymptotic Giant Branch (AGB)
star (e.g Bisterzo et al. 2011, Cristallo et al. 2015).


\begin{table*}
\caption[5]{\label{cnodr17noncal}
Revised CNO abundances derived from non-calibrated DR17 stellar parameters compared with
the DR17 CNO abundances in the last column.}
\begin{tabular}{l@{}c@{}c@{}c@{} | c@{}c@{}c@{} | c@{}c@{}c@{} | c@{}c@{}c@{} | l}
\noalign{\smallskip}
\hline
\noalign{\smallskip}
\hbox{ID}  & [C/Fe] & [N/Fe] & [O/Fe]   & [C/Fe] & [N/Fe] & [O/Fe] \\  
\noalign{\smallskip}
& \multicolumn{3}{c}{present work} &  \multicolumn{3}{c}{DR17} &   \\
\hline
\noalign{\smallskip}
 \noalign{\smallskip}
\noalign{\hrule}
 \noalign{\smallskip}
\noalign{\hrule}
\hline
b1 2M17153858-2759467	&	\phantom{-}	-0.60	&	0.40	&	0.35	&	-0.57	&	0.33	&	0.19	\\
b2 2M17173693-2806495	&	\phantom{-}	-0.20	&	0.00	&	0.40	&	-0.07	&	0.15	&	0.33	\\
b3 2M17250290-2800385	&	\phantom{-}	-0.05	&	0.10	&	0.35	&	0.09	&	0.15	&	0.32	\\
b4 2M17265563-2813558	&	\phantom{-}	-0.35	&	0.20	&	0.35	&	-0.29	&	0.29	&	0.30	\\
b5 2M17281191-2831393	&	\phantom{-}	-0.30	&	0.40	&	0.40	&	-0.18	&	0.23	&	0.30	\\
b6 2M17295481-2051262	&	\phantom{-}	-0.30	&	0.20	&	0.40	&	-0.07	&	0.04	&	0.35	\\
b7 2M17303581-2354453	&	\phantom{-}	-0.25	&	0.00	&	0.40	&	-0.06	&	0.17	&	0.35	\\
b8 2M17324257-2301417	&	\phantom{-}	+0.00	&	-0.10	&	0.35	&	0.11	&	0.11	&	0.33	\\
b9 2M17330695-2302130	&	\phantom{-} 0.00	&	0.00	&	0.35	&	0.18	&	-0.02	&	0.30	\\
b10 2M17344841-4540171	&	\phantom{-}	-0.30	&	0.20	&	0.35	&	-0.11	&	0.17	&	0.35	\\
b11 2M17351981-1948329	&	\phantom{-}	-0.10	&	0.10	&	0.40	&	0.00	&	0.04	&	0.17	\\
b12 2M17354093-1716200	&	\phantom{-}	-0.20	&	0.00	&	0.37	&	-0.03	&	0.13	&	0.32	\\
b13 2M17390801-2331379	&	\phantom{-}	-0.10	&	0.15	&	0.38	&	0.05	&	0.19	&	0.31	\\
b14 2M17392719-2310311	&	\phantom{-}	-0.10	&	0.10	&	0.38	&	0.02	&	0.18	&	0.26	\\
b15 2M17473299-2258254	&	\phantom{-}	-0.70	&	0.80	&	0.35	&	-0.49	&	0.45	&	0.29	\\
b16 2M17482995-2305299	&	\phantom{-}	-0.30	&	0.30	&	0.40	&	-0.43	&	0.54	&	0.34	\\
b17 2M17483633-2242483	&	\phantom{-}	-0.20	&	0.10	&	0.35	&	-0.08	&	0.10	&	0.19	\\
b18 2M17503263-3654102	&	\phantom{-}	-0.40	&	0.40	&	0.33	&	-0.17	&	0.22	&	0.33	\\
b19 2M17552744-3228019	&	\phantom{-}	-0.30	&	0.40	&	0.35	&	-0.18	&	0.20	&	0.32	\\
b20 2M18020063-1814495	&	\phantom{-}	-0.50	&	0.30	&	0.35	&	-0.42	&	0.24	&	0.24	\\
b21 2M18050452-3249149	&	\phantom{-}	-0.50	&	0.20	&	0.40	&	-0.29	&	0.26	&	0.32	\\
b22 2M18050663-3005419	&	\phantom{-}	-0.10	&	0.00	&	0.40	&	0.01	&	0.02	&	0.17	\\
b23 2M18065321-2524392	&	\phantom{-}	-0.20	&	0.20	&	0.38	&	-0.04	&	0.14	&	0.33	\\
b24 2M18104496-2719514	&	\phantom{-}	-0.10	&	0.10	&	0.35	&	-0.03	&	0.21	&	0.33	\\
b25 2M18125718-2732215	&	\phantom{-}	-0.22	&	0.20	&	0.40	&	-0.16	&	0.15	&	0.11	\\
b26 2M18200365-3224168	&	\phantom{-}	-0.35	&	0.20	&	0.32	&	-0.15	&	0.25	&	0.33	\\
b27 2M18500307-1427291	&	\phantom{-}	-0.30	&	0.20	&	0.38	&	-0.14	&	0.10	&	0.32	\\
c1 2M17173248-2518529	&	\phantom{-}	-0.25	&	0.20	&	0.38	&	-0.17	&	0.12	&	0.21	\\
c2 2M17285088-2855427	&	\phantom{-}	-0.45	&	0.40	&	0.40	&	-0.28	&	0.16	&	0.28	\\
c15 2M17291778-2602468	&	\phantom{-}	-0.20	&	0.30	&	0.38	&	-0.08	&	0.16	&	0.31	\\
c3 2M17301495-2337002	&	\phantom{-}	-0.25	&	0.20	&	0.40	&	-0.12	&	0.22	&	0.27	\\
c16 2M17310874-2956542	&	\phantom{-}	-0.40	&	0.20	&	0.36	&	-0.14	&	0.22	&	0.36	\\
c17 2M17382504-2424163	&	\phantom{-}	-0.20	&	0.30	&	0.40	&	-0.06	&	0.16	&	0.25	\\
c4 2M17453659-2309130	&	\phantom{-}	-0.30	&	0.30	&	0.40	&	-0.24	&	0.06	&	0.25	\\
c18 2M17511568-3249403	&	\phantom{-}	-0.20	&	0.00	&	0.38	&	-0.04	&	0.16	&	0.34	\\
c5 2M17532599-2053304	&	\phantom{-}	-0.25	&	0.20	&	0.40	&	-0.04	&	0.21	&	0.31	\\
c19 2M17552681-3334272	&	\phantom{-}	-0.30	&	0.00	&	0.40	&	-0.16	&	0.17	&	0.35	\\
c20 2M18005152-2916576	&	\phantom{-}	-0.40	&	0.20	&	0.40	&	-0.20	&	0.30	&	0.33	\\
c21 2M18010424-3126158	&	\phantom{-}	-0.25	&	0.00	&	0.38	&	-0.09	&	0.18	&	0.28	\\
c22 2M18042687-2928348	&	\phantom{-}	-0.50	&	0.30	&	0.40	&	-0.34	&	0.29	&	0.31	\\
c6 2M18044663-3132174	&	\phantom{-}	-0.15	&	0.00	&	0.40	&	0.02	&	0.14	&	0.31	\\
c23 2M18052388-2953056	&	\phantom{-}	-0.35	&	0.40	&	0.40	&	-0.45	&	0.34	&	0.38	\\
c7 2M18080306-3125381	&	\phantom{-}	-0.30	&	0.00	&	0.40	&	-0.09	&	0.05	&	0.40	\\
c24 2M18142265-0904155	&	\phantom{-}	-0.20	&	0.20	&	0.33	&	0.02	&	0.17	&	0.33	\\
c8 2M18195859-1912513	&	\phantom{-}	-0.40	&	0.40	&	0.40	&	-0.28	&	0.15	&	0.30	\\
c9 2M17190320-2857321	&	\phantom{-}	-0.30	&	0.20	&	0.70	&	-0.24	&	0.19	&	0.33	\\
c10 2M17224443-2343053	&	\phantom{-}	-0.35	&	0.20	&	0.37	&	-0.11	&	0.38	&	0.34	\\
c11 2M17292082-2126433	&	\phantom{-}	-0.60	&	0.60	&	0.38	&	-0.54	&	1.06	&	0.22	\\
c25 2M17293482-2741164	&	\phantom{-}	-0.50	&	0.30	&	0.40	&	-0.33	&	0.32	&	0.34	\\
c12 2M17323787-2023013	&	\phantom{-}	-0.15	&	0.20	&	0.40	&	0.04	&	0.13	&	0.33	\\
c13 2M17330730-2407378	&	\phantom{-}	-0.70	&	0.70	&	0.38	&	-0.73	&	0.30	&	0.30	\\
c26 2M17341796-3905103	&	\phantom{-}	-0.40	&	0.25	&	0.40	&	-0.07	&	0.20	&	0.34	\\
c27 2M17342067-3902066	&	\phantom{-}	-0.30	&	0.20	&	0.40	&	-0.14	&	0.29	&	0.45	\\
c28 2M17503065-2313234	&	\phantom{-}	-0.10	&	0.00	&	0.35	&	0.08	&	0.16	&	0.33	\\
c14 2M18023156-2834451	&	\phantom{-}	-0.05	&	0.10	&	0.40	&	0.01	&	0.07	&	0.18	\\
c29 2M18143710-2650147	&	\phantom{-}	-0.30	&	0.10	&	0.40	&	-0.14	&	0.27	&	0.34	\\
c30 2M18150516-2708486	&	\phantom{-}	-0.05	&	0.00	&	0.31	&	0.08	&	0.14	&	0.32	\\
c31 2M18344461-2415140	&	\phantom{-}	-0.45	&	0.40	&	0.40	&	-0.39	&	0.25	&	0.37	\\
\hline
\noalign{\smallskip}
\hline 
\end{tabular}
\end{table*}

\begin{table}
\scalefont{0.9}
\caption[4]{\label{revisedlight}
  Abundances from original APOGEE-ASPCAP derivations,
  and revised values of Ce, using the lines reported in
  Table \ref{linelist} for the 58 sample stars. For Ce abundances the two columns correspond to
 DR17 ASPCAP abundances, and revised values.}
\begin{tabular}{lrrrrrrrr}  
\noalign{\smallskip}
\hline
\noalign{\smallskip}
Star &  \multicolumn{2}{c}{[Ce/Fe]} &\\
\hline
\noalign{\smallskip}
\noalign{\smallskip}

\hbox{ID}&   \hbox{DR17} & revised \\
            & &  \\
 \noalign{\smallskip}
\noalign{\hrule}
\hline
 \noalign{\smallskip}
 2M17153858-2759467&     -0.16& 0.25  \\ 
 2M17173693-2806495&     -0.10& 0.20  \\ 
 2M17250290-2800385&     ---- & 0.20  \\ 
 2M17265563-2813558&    -0.20& 0.10   \\ 
 2M17281191-2831393&    -0.14& 0.20  \\
 2M17295481-2051262&    -0.02& -0.02  \\ 
 2M17303581-2354453&     ---- & 0.40   \\ 
 2M17324257-2301417&     ---- & ---  \\ 
 2M17330695-2302130&     ---- & 0.50  \\ 
 2M17344841-4540171&     ---- & 0.50  \\ 
 2M17351981-1948329&     ---- & 0.50  \\ 
 2M17354093-1716200&     ---- & 0.50  \\ 
 2M17390801-2331379&     ---- & 0.50  \\ 
 2M17392719-2310311&  ---- & 0.50  \\ 
 2M17473299-2258254&    -0.27& 0.30  \\  
 2M17482995-2305299&    -0.4 & 0.20  \\   
 2M17483633-2242483&     ---- & 0.50  \\  
 2M17503263-3654102&     ---- & 0.50  \\  
 2M17552744-3228019&    -0.15& 0.35  \\  
 2M18020063-1814495&    -0.08& 0.30  \\  
 2M18050452-3249149&     -0.11& 0.45  \\  
 2M18050663-3005419&     ---- & 0.40  \\  
 2M18065321-2524392&     ---- & 0.45  \\  
 2M18104496-2719514&     -0.17& 0.25  \\  
 2M18125718-2732215&     ---- & 0.30  \\  
 2M18200365-3224168&      0.07& 0.50  \\  
 2M18500307-1427291&      0.08& 0.45  \\  
 2M17173248-2518529&      0.03& 0.45  \\ 
 2M17285088-2855427&     ---- & 0.50  \\ 
 2M17291778-2602468&     ---- & 0.45  \\ 
 2M17301495-2337002&     ---- & 0.45  \\ 
 2M17310874-2956542&    -0.17& 0.30  \\ 
 2M17382504-2424163&     ---- & 0.10  \\ 
 2M17453659-2309130&    -0.40& -0.10  \\ 
 2M17511568-3249403&    -0.11& 0.40  \\ 
 2M17532599-2053304&     ---- & 0.40  \\ 
 2M17552681-3334272&     0.03& 0.35  \\ 
 2M18005152-2916576&   -0.12& 0.30  \\ 
 2M18010424-3126158&    ---- & 0.43  \\ 
 2M18042687-2928348&   -0.07& 0.20  \\ 
 2M18044663-3132174&    ---- & 0.33  \\ 
 2M18052388-2953056&   -0.29& 0.20  \\ 
 2M18080306-3125381&    0.15& 0.25  \\ 
 2M18142265-0904155&   -0.15& 0.30  \\ 
 2M18195859-1912513&   -0.17& 0.35  \\ 
 2M17190320-2857321&   -0.20& 0.32  \\ 
 2M17224443-2343053&    0.13& 0.55  \\ 
 2M17292082-2126433&   -0.12& 0.42  \\ 
 2M17293482-2741164&   -0.27& 0.30  \\ 
 2M17323787-2023013&   ---- & 0.42  \\ 
 2M17330730-2407378&   -0.18& 0.30  \\ 
 2M17341796-3905103&   -0.03& 0.20  \\ 
 2M17342067-3902066&   -0.18& 0.20  \\ 
 2M17503065-2313234&    ---- & 0.20  \\ 
 2M18023156-2834451&    ---- & 0.50  \\ 
 2M18143710-2650147&   -0.18& 0.20  \\ 
 2M18150516-2708486&    ---- & 0.25  \\ 
 2M18344461-2415140&   -0.28& 0.40  \\ 
 \noalign{\smallskip}
\hline

\noalign{\smallskip}
\hline 
\end{tabular}
\end{table}

\section{Conclusions}

We have selected 58 stars from the bulge sample of Queiroz et al. (2021) based on APOGEE, 
with characteristics to belong to the spheroidal, pressure-supported bulge. 
For this sample we have analysed lines of C, N, O, alpha-elements Mg, Si, Ca and neutron-capture element Ce. 
Our fits to the Mg, Si and Ca abundances from the original APOGEE
results using the ASPCAP software appeared to be in good agreement. 
We recomputed abundances for C, N, O, and Ce, assuming the spectroscopic
non-calibrated stellar parameters from APOGEE DR17. We report
differences in abundances of these elements.

We compare the abundances of these elements to literature data for bulge stars, and chemodynamical models by
Fria\c ca \& Barbuy (2017) - see also Barbuy et al. (2018a). These comparisons show compatibility of
the abundances of the sample stars with literature and models for Mg, Si, and Ca
in which a pressure supported component (spheroidal bulge) formed on a very short timescale (below 1 Gyr).
Similar results were suggested by other chemical evolution models (see Matteucci 2021 for a review),
and for stars with similar alpha element enhancements with asteroseismic ages (Miglio et al. 2021).

Nitrogen abundances show no exceptional enhancement for any of the sample stars, therefore there is
no evidence for these stars to be a result of multiple stellar populations in dissolved globular clusters.
The Ce abundance is enhanced in all stars, which would point out to a s-process origin of this element
already in the very early phases of chemical enrichment. This could have  been achieved with spinstars (e.g. Chiappini et al. 2011, Frischknecht et al. 2016), or alternatively due to mass transfer from a companion AGB star (e.g. Cristallo et al. 2015). This same conclusion was reached by
Barbuy et al. (2009, 2014, 2021b) regarding the globular cluster NGC~6522, but here, since all the present sample stars are enhanced
in Ce, all of them would have to be binaries with an AGB companion. Therefore the spinstars seem to be a more plausible explanation.

\section*{Acknowledgements}
RR acknowledges a CNPq master fellowship. TM acknowledges FAPESP postdoctoral fellowship no. 2018/03480-7. HE acknowledges a CAPES PhD fellowship. A.P.-V. and S.O.S. acknowledge the DGAPA-PAPIIT grant IA103122. SOS acknowledges a FAPESP PhD fellowship no. 2018/22044-3 and the support of the Deutsche Forschungsgemeinschaft (DFG, project number: 428473034).
BB acknowledges grants from FAPESP, CNPq and CAPES - Financial code 001.
J.G.F-T gratefully acknowledges the grant support provided by Proyecto Fondecyt Iniciaci\'on No. 11220340, and also from ANID Concurso de Fomento a la Vinculaci\'on Internacional para Instituciones de Investigaci\'on Regionales (Modalidad corta duraci\'on) Proyecto No. FOVI210020, and from the Joint Committee ESO-Government of Chile 2021 (ORP 023/2021).
D.G. gratefully acknowledges support from the ANID BASAL project ACE210002.
D.G. also acknowledges financial support from the Direcci\'on de Investigaci\'on y Desarrollo de
la Universidad de La Serena through the Programa de Incentivo a la Investigaci\'on de
Acad\'emicos (PIA-DIDULS).
The work of V.M.P. is supported by NOIRLab, which is managed by the Association of Universities for Research in Astronomy (AURA) under a cooperative agreement with the National Science Foundation.
MZ was funded by ANID FONDECYT Regular 1191505, ANID Millennium Institute of Astrophysics (MAS) under
grant ICN12\_009, the ANID BASAL Center for Astrophysics and Associated Technologies (CATA) through
grants AFB170002, ACE210002 and FB210003.
RR, BB, TM, HE, SOS, are part of the Brazilian Participation Group (BPG) in the Sloan Digital Sky Survey (SDSS), from the
Laborat\'orio Interinstitucional de e-Astronomia – LIneA, 
Brazil. Funding for the Sloan Digital Sky Survey IV has been provided by the Alfred P. Sloan Foundation, the U.S. Department of Energy Office of Science, and the Participating Institutions. SDSS acknowledges support and resources from the Center for High-Performance Computing at the University of Utah. The SDSS web site is www.sdss.org. 
SDSS is managed by the Astrophysical Research Consortium for the Participating Institutions of the SDSS Collaboration including the Brazilian Participation Group, the Carnegie Institution for Science, Carnegie Mellon University, Center for Astrophysics | Harvard \& Smithsonian (CfA), the Chilean Participation Group, the French Participation Group, Instituto de Astrofísica de Canarias, The Johns Hopkins University, Kavli Institute for the Physics and Mathematics of the Universe (IPMU) / University of Tokyo, the Korean Participation Group, Lawrence Berkeley National Laboratory, Leibniz Institut für Astrophysik Potsdam (AIP), Max-Planck-Institut für Astronomie (MPIA Heidelberg), Max-Planck-Institut für Astrophysik (MPA Garching), Max-Planck-Institut f\"ur Extraterrestrische Physik (MPE), National Astronomical Observatories of China, New Mexico State University, New York University, University of Notre Dame, Observatório Nacional / MCTI, The Ohio State University, Pennsylvania State University, Shanghai Astronomical Observatory, United Kingdom Participation Group, Universidad Nacional Autónoma de México, University of Arizona, University of Colorado Boulder, University of Oxford, University of Portsmouth, University of Utah, University of Virginia, University of Washington, University of Wisconsin, Vanderbilt University, and Yale University.
This work makes use of data from the European Space Agency (ESA) space mission Gaia.  The Gaia mission website is https://www.cosmos.esa.int/gaia. The Gaia archive website is https://archives.esac.esa.int/gaia.


\section*{Data Availability}

The observed data are from the APOGEE survey. The calculations and
fits to the lines can be requested to the authors. The code for
spectrum synthesis is available to be retrieved.








\end{document}